\documentclass[12pt]{article}
\usepackage{epsfig,amsmath,amssymb}
\usepackage{graphicx,psfrag} 
\renewcommand{\baselinestretch}{1.25}

\numberwithin{equation}{section}

\newcommand{\be}{\begin{equation}}
\newcommand{\ee}{\end{equation}}
\newcommand{\bea}{\begin{eqnarray}}
\newcommand{\eea}{\end{eqnarray}}
\newcommand{\bA}{\begin{array}}
\newcommand{\eA}{\end{array}}
\newcommand{\bc}{\begin{center}}
\newcommand{\ec}{\end{center}}
\newcommand{\al}{\alpha}

\newcommand{\ra}{\rightarrow}
\newcommand{\del}{\partial}

\newcommand{\ie}{{\it i.e.}}
\newcommand{\eg}{{\it e.g.}}

\begin{document}

\begin{titlepage}
\vspace{30mm}
\bc
	\hfill 
	\\         [22mm]

	{\Huge On $AdS_2$ holography from redux,\\ [2mm]
        renormalization group flows and $c$-functions}   
	\vspace{16mm}
	
	{\large Kedar S. Kolekar,\ \ K. Narayan}\\
	\vspace{3mm}
	{\small \it Chennai Mathematical Institute, \\ }
	{\small \it SIPCOT IT Park, Siruseri 603103, India.\\ }
	
\ec
\medskip
\vspace{30mm}
	
\begin{abstract}
Extremal black branes upon compactification in the near horizon throat
region are known to give rise to $AdS_2$ dilaton-gravity-matter
theories. Away from the throat region, the background has nontrivial
profile. We interpret this as holographic renormalization group flow
in the 2-dim dilaton-gravity-matter theories arising from dimensional
reduction of the higher dimensional theories here. The null energy
conditions allow us to formulate a holographic c-function in terms of
the 2-dim dilaton for which we argue a c-theorem subject to
appropriate boundary conditions which amount to restrictions on the
ultraviolet theories containing these extremal branes. At the infrared
$AdS_2$ fixed point, the c-function becomes the extremal black brane
entropy.  We discuss the behaviour of this inherited c-function in
various explicit examples, in particular compactified nonconformal
branes, and compare it with other discussions of holographic
c-functions. We also adapt the holographic renormalization group
formulated in terms of radial Hamiltonian flow to 2-dim
dilaton-gravity-scalar theories, which while not Wilsonian, gives
qualitative insight into the flow equations and $\beta$-functions.
\end{abstract}

\end{titlepage}

{\tiny 
\begin{tableofcontents}
\end{tableofcontents}
}

\section{Introduction}

2-dimensional dilaton gravity is an interesting and relatively simple
playground for various physical questions, arising generically from
dimensional reduction of higher dimensional gravitational theories,
as is well known. In particular the near horizon geometry of extremal
black holes and branes in these theories is of the form $AdS_2\times
X$: compactifying the transverse space $X$ gives rise to effective
2-dim dilaton-gravity theories with $AdS_2$ arising as an attractor
point with constant dilaton (which controls the size of $X$).  In
recent years, 2-dim $AdS_2$ dilaton-gravity theories have been under
extensive investigation \cite{Almheiri:2014cka,Maldacena:2016upp,
  Jensen:2016pah,Engelsoy:2016xyb,Almheiri:2016fws}.  In these $AdS_2$
dilaton-gravity theories, the varying dilaton leads to the breaking of
the isometries of $AdS_2$, which amounts to breaking of local boundary
time reparametrizations (modulo global $SL(2)$ symmetries). The
leading effects describing such nearly $AdS_2$ theories are captured
universally by a Schwarzian derivative action governing boundary time
reparametrizations modulo $SL(2)$, from the leading nonconstant
dilaton behaviour, consistent with the absence of finite energy
excitations in $AdS_2$ \cite{Maldacena:1998uz,Sen:2008vm}. There are
parallels with recent investigations of the SYK model
\cite{Sachdev:1992fk,Kitaev-talks:2015,Polchinski:2016xgd,Maldacena:2016hyu,Kitaev:2017awl},
a quantum mechanical model of interacting fermions, which exhibits
approximate conformal symmetry at low energies, with the leading
departures governed by a Schwarzian action. See
\eg\ \cite{Sarosi:2017ykf,Rosenhaus:2018dtp} for a review of recent
developments.

Away from the $AdS_2$ throat region, the 2-dim theory exhibits
nontrivial evolution and it is interesting to ask if this can be
interpreted as a holographic renormalization group flow. There is a
long and rich history of formulating versions of the renormalization
group in the holographic context, beginning with
\eg\ \cite{Akhmedov:1998vf,Alvarez:1998wr,Girardello:1998pd,Distler:1998gb,Balasubramanian:1999jd,Skenderis:1999mm}. The
central feature here is the correspondence between the radial
coordinate in the bulk spacetime and the energy scale in the boundary
field theory \cite{Susskind:1998dq,Peet:1998wn}: evolution towards the
interior in the bulk corresponds to flowing to lower energies in the
boundary theory. In
\cite{deBoer:1999tgo,Verlinde:1999xm,deBoer:2000cz}, the holographic
renormalization group flow was formulated in terms of a radial
Hamiltonian evolution, which while not Wilsonian, provides useful
insights into the structure of the RG flow and $\beta$-functions.  The
striking Zamolodchikov c-theorem \cite{Zamolodchikov:1986gt} argues
that for 2-dim quantum field theories, there exists a positive
definite function of couplings that is monotonically decreasing along
RG flows, stationary at fixed points and equals the central charge of
the corresponding CFT.  Holographic versions of c-theorems were
discussed in
\cite{Freedman:1999gp,Bousso:1999xy,Sahakian:1999bd,Goldstein:2005rr}:
the monotonicity of the associated c-functions stems ultimately from
the null energy conditions which in turn encode the focussing property
of null geodesic congruences. Wilsonian versions of the holographic
renormalization group were formulated in
\cite{Heemskerk:2010hk,Faulkner:2010jy}. Various versions of
c-theorems have also been motivated by studies of entanglement
entropy: a recent review is \cite{Nishioka:2018khk}.

In this paper we formulate a version of holographic renormalization
group flows restricting attention to cases where the far infrared bulk
geometry acquires an $AdS_2$ throat, as occurs for extremal black
holes and branes. Further restricting to cases where the transverse
space is sufficiently symmetric, as \eg\ for extremal branes that
enjoy space/time translational symmetry and spatial rotational
symmetry, the transverse part of the bulk spacetime evolves only in
terms of its overall size (or warping). Then the essential flow
becomes 2-dimensional in the bulk and can be isolated by dimensional
reduction to appropriate 2-dim dilaton-gravity-matter theories.  (The
effect of the gauge fields that gave rise to charge is mimicked by an
appropriate potential for the dilaton and other scalars).  This
investigation was motivated by \cite{Kolekar:2018sba} where the
dimensional reduction of extremal black branes in 4-dim (relativistic)
Einstein-Maxwell and (nonrelativistic) hyperscaling violating
Lifshitz, hvLif, theories was studied to $AdS_2$
dilaton-gravity(-scalar) theories, as well as the leading departures
from $AdS_2$\ (similar embedddings have been studied recently in
\cite{Castro:2014ima,Cvetic:2016eiv,Das:2017pif,Taylor:2017dly,
  Gaikwad:2018dfc,Nayak:2018qej,Cadoni:2017dma,Li:2018omr,Castro:2018ffi}).
Since the bulk flow to the infrared $AdS_2$ is essentially
2-dimensional, our formulation does not really distinguish whether the
higher dimensional completion is relativistic or nonrelativistic. In
the far infrared, the $AdS_2$ fixed point is the very near horizon
region of the corresponding compactified extremal black brane and so
it is reasonable to take the central charge of the dual $CFT_1$ to be
the extremal entropy of the black brane, which is the number of
underlying microstates. The extremal entropy is given by the
transverse area\ ${V_{D-2} \Phi_h^2\over 4G_D} = {\Phi_h^2\over
  4G_2}$\ where the 2-dim dilaton $\Phi^2=g_{ii}^{(D-2)/2}$ controls
the size of the transverse space. This suggests formulating a
holographic c-function\ ${\cal C}(u) = {\Phi(u)^2\over 4G_2}$\ away
from the $AdS_2$ region where the bulk has a nontrivial flow. We argue
that this c-function monotonically decreases under flow towards the
interior (infrared) and satisfies a c-theorem that follows from the
null energy conditions and requiring appropriate boundary conditions
(that the $AdS_2$ throat arises in the nonrelativistic hvLif family
above; this is a fairly broad family that includes $AdS_D$,
nonconformal branes and so on, but is otherwise not ``too generic'').
In addition ${\cal C}(u) \ra S_{BH}$ at the infrared $AdS_2$ fixed
point, which then fixes the precise form of ${\cal C}$. This dilatonic
c-function has also previously been discussed in
\cite{Goldstein:2005rr} in the context of nonsupersymmetric 4-dim
black hole attractors \cite{Goldstein:2005hq}, which we were in part
motivated by: the present context and discussion is however different
in detail as will be clear in what follows.

In sec.~\ref{neccfn}, we study the null energy conditions and discuss
this c-function, with some explicit analysis in the phase diagram of
nonconformal D2-branes (sec.\,\ref{cfnD2M2}) and nonconformal D4-branes
(sec.\,\ref{cfnM5D4}). In sec.\,\ref{Entrcfn}, we compare this dilatonic
c-function with the entropic c-function that has been discussed in the
context of entanglement. While the entropic c-function $c_E$ scales as
the number of local degrees of freedom (this is also the scaling of
the c-function in \cite{Freedman:1999gp}), the dilatonic c-function
above is extensive: it scales as the transverse area. Loosely
speaking, ${\cal C}\sim c_E V_{d_i} w^{d_i}$ where the $AdS_2$ throat
arises after compactification from $AdS_2\times X^{d_i}$.

In sec.~\ref{dVV2d}, we adapt the holographic RG formulation of de
Boer, Verlinde, Verlinde \cite{deBoer:1999tgo} to 2-dim
dilaton-gravity-scalar theories. In particular, we obtain RG flow
equations and $\beta$ functions for the (scalar) couplings in the
1-dim boundary theory in a derivative expansion.  Using this, we
compute $\beta$-functions for 2-dim bulk theories arising from
reductions of conformal and nonconformal branes. This suggests that it
is not consistent to place the $AdS_2$ throat in a bulk region which
exhibits nontrivial RG flow (\ie\ the $AdS_2$ throat needs to lie
within the bulk region corresponding to the RG fixed point), and
resolves a concern about apparently massless perturbations found in
\cite{Kolekar:2018sba}.  This is not Wilsonian: it would be
interesting to adapt the holographic Wilsonian RG of
\cite{Heemskerk:2010hk,Faulkner:2010jy} to these 2-dim theories and we
leave this for future work.
Sec.-6 contains a Discussion and the Appendices contain various
technical details.

\section{Reviewing $AdS_2$ dilaton gravity from higher dimensional
  redux}\label{Review}


In this section we review our discussion in \cite{Kolekar:2018sba},
where we studied $2$-dim $AdS_2$ dilaton-gravity matter theories
arising from dimensional reduction of extremal charged black branes in
Einstein-Maxwell and hyperscaling violating Lifshitz theories in
$4$-dimensions. The $AdS_2$ throat in the near horizon geometry of
these extremal black branes manifests as the constant dilaton,
constant scalar field $AdS_2$ background in the effective 2-dim
theory. The leading departures away from $AdS_2$ are governed by the
Schwarzian derivative action, arising from the Gibbons-Hawking term
linear in the dilaton perturbation, consistent with earlier results.

\noindent\underline{Einstein-Maxwell theory}:\ \ 
Einstein-Maxwell theory with negative cosmological constant in
4-dimensions admits electrically and magnetically charged black
brane solutions (which asymptote to $AdS_4$). A review of various
features of relevance here and later is \cite{Hartnoll:2016apf}.
The action is
\begin{equation}\label{cbb4daction}
S=\int d^4x\sqrt{-g^{(4)}}\left[\frac{1}{16\pi G_4}\Big(\mathcal{R}^{(4)}-2\Lambda\Big)-\frac{1}{4}F_{MN}F^{MN}\right]\ , \qquad \Lambda=-\frac{3}{R^2}\ ,
\end{equation}
where $R$ is the $AdS_4$ radius. The electric black brane metric is
\begin{equation}\label{cbb4dsolne}
ds^2=-\frac{r^2f(r)}{R^2}dt^2+\frac{R^2}{r^2f(r)}dr^2+\frac{r^2}{R^2}(dx^2+dy^2)\ , \qquad f(r)=1-\left(\frac{r_0}{r}\right)^3+\frac{Q_e^2}{r^4}\Big(1-\frac{r}{r_0}\Big)
\end{equation}
and the gauge field is $A_t=\mu (1-\frac{r_0}{r})$, where $r_0$ is the
horizon, $r\ra\infty$ is the boundary and the charge parameter $Q_e$
is related to the chemical potential $\mu$ as\
$\mu={Q_e\over 2\sqrt{\pi G_4}\,Rr_0}$\ \cite{Hartnoll:2016apf}.

In the extremal limit, the Hawking temperature $T=\frac{3r_0}{4\pi
  R^2}(1-\frac{Q_e^2}{3r_0^4})$ vanishes giving $Q_e^4=r_0^4$ and the
near horizon geometry of the black brane becomes
$AdS_2\times\mathbb{R}^2$. This $AdS_2$ throat is well defined in the
regime $\frac{r-r_0}{R}\gg 1$ and $\frac{r-r_0}{r_0}\ll 1$ and is well
separated from the asymptotic $AdS_4$ boundary at $r\sim r_c \gg r_0$
if $\frac{r-r_0}{r_c}\ll 1$. The Bekenstein-Hawking entropy\ $S_{BH}=
\frac{V_2}{4G_4}\frac{r_0^2}{R^2}=\frac{V_2}{4G_4}\frac{Q_e}{\sqrt{3}R^2}$,
with $V_2=\int dxdy$, amounts to finite entropy density for noncompact
branes.

We compactify the transverse 2-dim space as a torus $T^2$ with a
KK-reduction ansatz\ $ds^2=g^{(2)}_{\mu\nu}dx^{\mu}dx^{\nu} +\Phi^2(dx^2+dy^2)$
for the metric. Then performing a Weyl transformation
$g_{\mu\nu}=\Phi g^{(2)}_{\mu\nu}$ of the $2$-dim metric to
absorb the kinetic term for the dilaton $\Phi$ in the Ricci scalar, the
action \eqref{cbb4daction} reduces to\
$S=\int d^2x\sqrt{-g}\, \Big[\frac{1}{16\pi G_2}(\Phi^2\mathcal{R}-2\Lambda\Phi)-\frac{V_2\Phi^3}{4}F_{\mu\nu}F^{\mu\nu}\Big]$,\ where $G_2=G_4/V_2$\
for the electric brane solution. We solve for the gauge field in terms of
the dilaton
\ie\ $F^{\mu\nu}=\frac{Q_e}{2\sqrt{\pi G_4}\,R^3}\frac{1}{\sqrt{-g}\,\Phi^3}\,\varepsilon^{\mu\nu}$, with $\varepsilon^{tr}=1$, $\varepsilon_{\mu\nu}=g_{\mu\alpha}g_{\nu\beta}\varepsilon^{\alpha\beta}$ and substitute $F^{\mu\nu}$ in this
dilaton-gravity-Maxwell action (with a sign change for electric
branes stemming from $g_{tt}$). This gives an equivalent dilaton-gravity
theory with a dilaton potential, which now encodes the gauge field profile,
\be\label{cbb2dequivaction}
S=\frac{1}{16\pi G_2}\int d^2 x\sqrt{-g}\,
\Big(\Phi^2\mathcal{R}- U(\Phi)\Big)\ , \qquad U(\Phi)=2\Lambda\Phi+\frac{2Q_e^2}{R^6\Phi^3}\ .
\ee
This equivalent dilaton-gravity theory admits an $AdS_2$ solution with
a constant dilaton, which is just the near horizon $AdS_2$ throat region
of the 4-dim extremal black brane.

Turning on perturbations to the dilaton and the metric, we see that
the quadratic part of the bulk action gives coupled linearized
equations governing these perturbations, while the leading correction
comes from the Gibbons-Hawking term. This leading term is linear in
the dilaton perturbation $\tilde{\phi}$ and gives the Schwarzian
derivative (Euclidean) action upon expanding the extrinsic curvature
around $AdS_2$ background
\begin{equation}\label{SchwEntrz=1}
S_{GH}^{(1)}=-\frac{2\Phi_h^2}{8\pi G_2}\int d\tau\sqrt{\gamma}\
{\tilde\phi}\, K\ \longrightarrow\
-\frac{\Phi_h^2}{4\pi G_2}\int du\, \phi_r(u)\, \{\tau(u),u\}\ ,
\end{equation}
where $\Phi_h=\frac{r_0}{R}$. In evaluating the last term, we take the
boundary of $AdS_2$ as a slightly deformed curve $(\tau(u),\rho(u))$
parametrized by the boundary
coordinate $u$, and define ${\tilde\phi}={\phi_r(u)\over\epsilon}$\,,\
as discussed in \cite{Maldacena:2016upp}\ (reviewed in
\cite{Sarosi:2017ykf}). The Schwarzian derivative action is
$Sch(\tau(u),u)=\{\tau(u),u\}={\tau'''\over \tau'}-{3\over 2}
({\tau''\over \tau'})^2$.

\noindent\underline{Charged hvLif black branes}:\ \ 
Charged hyperscaling violating Lifshitz (hvLif) black branes in
4-dimensions arise in Einstein-Maxwell-scalar theories with
an additional $U(1)$ gauge field, where one $U(1)$ gauge field and the
scalar source the nonrelativistic background while the other $U(1)$
gauge field gives charge to the black brane \cite{Tarrio:2011de,Alishahiha:2012qu,Bueno:2012sd}. The action is
\begin{equation}\label{chbb4daction}
S=\int d^4x\sqrt{-g^{(4)}}\Big[\frac{1}{16\pi G_4}\Big(\mathcal{R}^{(4)}-\frac{1}{2}\partial_M\Psi\partial^M\Psi+V(\Psi)-\frac{Z_1}{4}(F_1)^2\Big)-\frac{Z_2}{4}(F_2)^2\Big],
\end{equation}
where $(F_i)^2=F_{i\,MN}F^{MN}$. The scalar dependent couplings and the scalar potential are $Z_1=e^{\lambda_1\Psi}$, $Z_2=e^{\lambda_2\Psi}$ and $V(\Psi)=V_0 e^{\gamma\Psi}$. The charged hvLif black brane metric is
\begin{eqnarray}
ds^2&=&\Big(\frac{r}{r_{hv}}\Big)^{-\theta}\Big[-\frac{r^{2z}f(r)}{R^{2z}}dt^2+\frac{R^2}{r^2f(r)}dr^2+\frac{r^2}{R^2}(dx^2+dy^2)\Big]\ , \nonumber\\[2pt]
f(r)&=&1-\left(\frac{r_0}{r}\right)^{2+z-\theta}+\frac{Q^2}{r^{2(1+z-\theta)}}\Big(1-\Big(\frac{r}{r_0}\Big)^{z-\theta}\Big)\ . \label{chargedhvLifmetric}
\end{eqnarray}
These are asymptotically hvLif (see \eg\ \cite{Hartnoll:2016apf}
on hvLif backgrounds). The scalar and gauge fields are
\begin{eqnarray}
e^{\Psi}&=&e^{\Psi_0}\Big(\frac{r_{hv}\,r}{R^2}\Big)^{\sqrt{(2-\theta)(2z-2-\theta)}}, \quad F_{1rt}=\sqrt{2(z-1)(2+z-\theta)}\,e^{-\frac{\lambda_1\Psi_0}{2}}\,r_{hv}^2\,
R^{\theta-z-4}\, r^{1+z-\theta}, \nonumber \\
F_{2rt}&=&\frac{Q\sqrt{2(2-\theta)(z-\theta)}\,e^{-\frac{\lambda_2\Psi_0}{2}}}{4\sqrt{\pi G_4}}\,R^{z-\theta-2}\,r_{hv}^{-z+\theta+1}\,r^{-(1+z-\theta)}, \label{4dchargedhvLiffields}
\end{eqnarray}
with $V_0=\frac{(2+z-\theta)(1+z-\theta)\,e^{-\gamma\Psi_0}}{R^{2-2\theta}\,r_{hv}^{2\theta}}$, $\gamma=\frac{\theta}{\sqrt{(2-\theta)(2z-2-\theta)}}$, $\lambda_1=\frac{-4+\theta}{\sqrt{(2-\theta)(2z-2-\theta)}}$, $\lambda_2=\sqrt{\frac{2z-2-\theta}{2-\theta}}$, where $z$ and $\theta$ are Lifshitz and hyperscaling violation exponents respectively. $r_{hv}$ is the hyperscaling violating scale arising in the conformal
factor in the metric, and the charge parameter $Q$ has dimensions of
$r^{1+z-\theta}$\ (which is equivalent to absorbing factors of $r_{hv}, R$
into $Q$). The null energy conditions and the regularity of $A_{2\,t}$ as
$r\rightarrow\infty$, so that hvLif boundary is not ruined by $A_{2\,t}$,
give constraints on the exponents $z$, $\theta$,
\be\label{chbbzthetarange}
z\geq 1\ ,\qquad\quad 2z-2-\theta\geq 0\ ,\qquad\quad 2-\theta\geq 0\ .
\ee

In the extremal limit, the temperature
$T=\frac{(2+z-\theta)r_0^z}{4\pi R^{z+1}}(1-\frac{(z-\theta)Q^2
  r_0^{-2(1+z-\theta)}}{(2+z-\theta)})$ vanishes giving
$Q^2=\frac{(2+z-\theta)}{(z-\theta)}r_0^{2(1+z-\theta)}$ and an
$AdS_2\times\mathbb{R}^2$ near horizon geometry. The $AdS_2$ throat,
well defined if $\frac{r-r_0}{R}\gg 1$, $\frac{r-r_0}{r_0}\ll 1$, is
well separated from the asymptotic hvLif boundary taken at $r\sim
r_{hv} \gg r_0$ if $\frac{r-r_0}{r_{hv}}\ll 1$.

As in the relativistic case earlier, a $T^2$-redux followed by a Weyl
transformation and replacing the gauge fields $F_1^{\mu\nu}$, $F_2^{\mu\nu}$
in terms of the dilaton and the scalar field $\Psi$ gives an equivalent
dilaton-gravity-scalar action with an effective interaction potential
$U(\Phi,\Psi)$ for the dilaton and the scalar field,
\bea\label{chbb2deffactionWeyl}
&& S=\frac{1}{16\pi G_2}\int d^2 x\sqrt{-g}\ \Big(\Phi^2\mathcal{R}-\frac{\Phi^2}{2}(\partial\Psi)^2-U(\Phi,\Psi)\Big), \qquad \\[2pt]
&& U(\Phi,\Psi)=-\frac{(2+z-\theta)(1+z-\theta)}{R^{2-2\theta}r_{hv}^{2\theta}}e^{\gamma(\Psi-\Psi_0)}\,\Phi \qquad\qquad \nonumber\\
&& \qquad\qquad +\, \frac{1}{\Phi^3}\Big(\frac{(z-1)(2+z-\theta)r_{hv}^{2\theta-4}R^{2-2\theta}}{e^{\lambda_1(\Psi-\Psi_0)}}+\frac{(2-\theta)(z-\theta)Q^2 r_{hv}^{2z-2}R^{-4z-2+2\theta}}{e^{\lambda_2(\Psi-\Psi_0)}}\Big)\ . \nonumber
\eea
This admits an $AdS_2$ background with constant dilaton and scalar field,
which is the near horizon $AdS_2$ throat of the 4-dim extremal hvLif black
brane.

Turning on a perturbation to the dilaton $\Phi=\Phi_h(1+{\tilde\phi})$
(and the scalar field and metric), we see that the quadratic part of
the bulk action gives coupled linearized equations governing these
perturbations. Using conformal gauge, $AdS_2$ is\
$ds^2=\frac{4L^2}{(x^+-x^-)^2}(-dx^+dx^-)$, and the (decoupled)
linearized equation for the scalar field perturbation\
$(2z-2-\theta)\partial_+\partial_-\zeta+2(z-1){2\over (x^+-x^-)^2}\zeta=0$\
shows that $\zeta$ is massive generically and the
$AdS_2$ background is stable for $z$, $\theta$ satisfying the energy
conditions \eqref{chbbzthetarange}. However, for $z=1$, $\theta\neq0$,
$\zeta$ is massless and suggests that the linear stability analysis is
insufficient to determine the stabilty of the $AdS_2$ attractor. An
example where this case arises is the reduction of $D2$-branes on $S^6$
which gives a 4-dim hvLif theory with $z=1,\ \theta=-\frac{1}{3}$\
(see Appendix \ref{z=1nonzerotheta}).

The leading correction to $AdS_2$ comes from the Gibbons-Hawking term
linear in the dilaton perturbation, which gives the Schwarzian
derivative action as mentioned earlier. We expand the total (Euclidean)
action as $I=I_0+I_1+I_2+\dots$ in perturbations. 
The background action $I_0=-\frac{\Phi_h^2}{16\pi G_2}
(\int d^2x \sqrt{g}\,\mathcal{R} +2\int_{{\tiny bndry}}
\sqrt{\gamma}\,K)$ gives the extremal entropy\
$S_{BH}=\frac{\Phi_h^2}{4G_2}$\ where $\Phi_h^2=\frac{r_0^{2-\theta}}{R^2 r_{hv}^{-\theta}}$\,. $I_1$ is linear in perturbations and when evaluated on the
$AdS_2$ background with constant dilaton $\Phi_h$ and scalar $\Psi_h$ gives
\begin{equation}
I_1=-\frac{2\Phi_h^2}{16\pi G_2}\int d^2x\,\sqrt{g}\,
\tilde{\phi}\,\Big(\mathcal{R}+\frac{2}{L^2}\Big)-\frac{2\Phi_h^2}{8\pi G_2}\int_{bndry}\sqrt{\gamma}\,\tilde{\phi}\,K\ ,
\end{equation}
which is the Jackiw-Teitelboim \cite{Jackiw:1984je,Teitelboim:1983ux} theory.
The entropy coefficients arise from the way the perturbations have been
defined. It is worth noting that the appearance of the Schwarzian from
the Gibbons-Hawking term applies to the JT model as in these cases
above. In what follows, we study more general 2-dim
dilaton-gravity-matter theories: in such cases, the appearance of the
Schwarzian follows from a more general argument
\cite{Cvetic:2016eiv,Castro:2018ffi}, related to a conformal anomaly
arising after properly renormalizing the on-shell action.


\section{The 2-dim theory and the attractor conditions}

We consider a general gravity-scalar action in $D$ dimensions
\begin{equation}\label{Ddimgsaction}
S=\frac{1}{16\pi G_D}\int d^Dx\sqrt{-g^{(D)}}\ \Big(\mathcal{R}^{(D)}
-\frac{h_{IJ}}{2}\partial_M\Psi^I\partial^M\Psi^J-V \Big)\ ,
\end{equation}
where $h_{IJ}(\Psi^I)$ is a positive definite, symmetric metric
controlling the kinetic terms of the scalars $\Psi^I$ and
$V=V(\Psi^I,g)$ is a potential for the scalars $\Psi^I$ which also
contains metric data (\ie\ $V$ is not simply a scalar potential). Such
an effective action arises from theories with gravity, scalars and
gauge fields after the gauge fields have been replaced with their
background profiles (and changing the signs of the $F^2$ terms for
electric profiles): we have seen examples of this sort arising in
Einstein-Maxwell and Einstein-Maxwell-dilaton theories in the previous
section. For instance, in the Einstein-Maxwell case with no scalars,
the term $\int \sqrt{-g} F^2$ gives\ $\del_r (\sqrt{-g} F^{tr})=0$ for
electric branes: using this $F^{tr}$-profile gives the
term\ ${g_{tt}g_{rr}\over g}F_0^2$ thus leading to the effective
potential $V=-V_0(\Psi^I) + {1\over g_{xx}^{D-2}}V_2(\Psi^I)$, with
$V=-V_0(\Psi^I)$ arising from the cosmological constant term in the
original theory. The sign of the $V_2(\Psi^I)$-term is fixed by
requiring that the gravity-scalar equations are identical with those
of the original theory (See Appendix \ref{DdimgsVeff} for
details). Note that this sign is also consistent with
electric-magnetic duality (for magnetic branes, the $F^2$ term does
not contain a minus sign which only arises from $g_{tt}$ for electric
branes).  It is worth noting that the gauge fields have not really
been ``integrated out'' and so these gravity-scalar theories are best
regarded as equivalent only for certain (classical or semiclassical)
purposes as will be clear in what follows.

We now look for 2-dim theories obtained by dimensional reduction of
the above theories on a torus $T^{D-2}$ with the ansatz
\begin{equation}\label{DdimKKmetric}
ds^2=g^{(2)}_{\mu\nu}dx^{\mu}dx^{\nu}+\Phi^{\frac{4}{D-2}}\sum_{i=1}^{D-2}dx_i^2\ ,
\qquad\qquad g_{xx}^{(D)}\equiv \Phi^{\frac{4}{D-2}}\ .
\end{equation}
This gives the 2-dim action
\begin{equation}\label{2dimgsactionnoWeyl}
S=\frac{1}{16\pi G_2}\int d^2 x\sqrt{-g^{(2)}}\ \Phi^2\,\Big(\mathcal{R}^{(2)}+\ \frac{D-3}{D-2}\frac{(\nabla_{(2)}\Phi^2)^2}{\Phi^4}-\ \frac{2\nabla_{(2)}^2\Phi^2}{\Phi^2}-\ \frac{h_{IJ}}{2}\partial_{\mu}\Psi^I\partial^{\mu}\Psi^J-\ V \Big)\ ,
\end{equation}
where $\nabla_{(2)\,\mu}$ is a covariant derivative w.r.t. $g^{(2)}_{\mu\nu}$.\
Now performing a Weyl transformation\
$g_{\mu\nu}=\Phi^{\frac{2(D-3)}{(D-2)}}g^{(2)}_{\mu\nu}$\ absorbs the kinetic
term\footnote{\label{2dimricciWeyl}
	Using the covariant derivative $\nabla_{\mu}$ w.r.t. $g_{\mu\nu}$, and
	\bea
	\nabla_{(2)}^2\Phi^2=\Phi^{\frac{2(D-3)}{(D-2)}}\nabla^2\Phi^2 ,\ \ \
	(\nabla_{(2)}\Phi^2)^2=\Phi^{\frac{2(D-3)}{(D-2)}}(\nabla\Phi^2)^2 ,\ \ \
	\mathcal{R}^{(2)}=\Phi^{\frac{2(D-3)}{(D-2)}}\left[\mathcal{R}-\frac{D-3}{D-2}\left(\frac{(\nabla\Phi^2)^2}{\Phi^4}-\frac{\nabla^2\Phi^2}{\Phi^2}\right)\right] .
	\nonumber
	\eea
}
for $\Phi$ in $\mathcal{R}$. The 2-dim action then becomes
\begin{equation}\label{2dimgsaction}
S=\frac{1}{16\pi G_2}\int d^2x\sqrt{-g}\Big(\Phi^2\mathcal{R}-\frac{\Phi^2}{2}h_{IJ}\partial_{\mu}\Psi^I\partial^{\mu}\Psi^J-U(\Phi,\Psi^I) \Big)\ , \qquad U(\Phi,\Psi^I)=V\Phi^{\frac{2}{D-2}}\ .
\end{equation}
We have suppressed a total derivative term\
$\int d^2x\sqrt{-g}\,[-\frac{(D-1)}{(D-2)}\nabla^2\Phi^2]$\ which cancels
with a corresponding term arising from the reduction of the 
Gibbons-Hawking boundary term.

Our choice in \eqref{DdimKKmetric} of the 2-dim dilaton\
$\Phi^2=g_{xx}^{(D-2)/2}$ implies that the area of the transverse
space is given by $\Phi^2$:\ also this choice leads to $\int (\Phi^2
{\cal R} + \ldots)$ uniformly in the Einstein term of the 2-dim action
for any higher dimensional theory.

The 2-dim equations of motion then become
\begin{eqnarray}\label{2dimgseoms}
g_{\mu\nu}\nabla^2\Phi^2-\nabla_{\mu}\nabla_{\nu}\Phi^2+\frac{g_{\mu\nu}}{2}\Big(\frac{\Phi^2}{2}h_{IJ}\partial_{\mu}\Psi^I\partial^{\mu}\Psi^J+U\Big)-\frac{\Phi^2}{2}h_{IJ}\partial_{\mu}\Psi^I\partial_{\nu}\Psi^J&=&0\ , \nonumber \\
\mathcal{R}-\frac{h_{IJ}}{2}\partial_{\mu}\Psi^I\partial^{\mu}\Psi^J-\frac{\partial U}{\partial(\Phi^2)}&=&0\ , \nonumber \\
\frac{1}{\sqrt{-g}}\partial_{\mu}(\sqrt{-g}\Phi^2 h_{IJ}\partial^{\mu}\Psi^J)-\frac{\partial U}{\partial\Psi^I}&=&0\ .
\end{eqnarray}
These equations admit an $AdS_2$ critical point with constant scalars
and dilaton: we have $\Phi,\ \Psi^I=const$, and ${\cal R}=-{2\over L^2}$\
(with $L$ the $AdS_2$ scale) which implies
\begin{equation}\label{AdS2conditions}
U_h=0\ , \qquad
\frac{\partial U}{\partial(\Phi^2)}\Big\rvert_h=\frac{-2}{L^2}\ ,
\qquad \frac{\partial U}{\partial\Psi^I}\Big\rvert_h=0\ ,
\end{equation}
from the first, second and third equations respectively; the subscript
$h$ denotes that the quantity is evaluated at the $AdS_2$ background (which is
the near horizon throat region of the higher dimensional extremal brane). While we focus in this paper on pure $AdS_2$ backgrounds with constant dilaton and constant scalars, the field equations \eqref{2dimgseoms} admit other solutions including a 2-dim black hole (which is locally $AdS_2$) where the conditions in \eqref{AdS2conditions} are modified (see \eg\ \cite{Castro:2018ffi}).
Turning on perturbations,
\begin{equation}
\Phi=\Phi_h+\phi\ , \qquad \Psi^I=\Psi^I_h+\psi^I\ , \qquad \omega=\omega_h+\Omega\ ,
\end{equation}
where $ds^2=e^{2\omega}(-dx^+ dx^-)$ (conformal gauge), the linearized field equations for these perturbations are
\begin{eqnarray}\label{2d-fluctEOM}
	&& \hspace{94mm} \partial_+\partial_-\phi + \frac{2\phi}{(x^+ - x^-)^2} = 0\ , \nonumber \\
	&& \hspace{12mm} (h_{IJ}|_h)\partial_+\partial_-\psi^J + \frac{L^2}{(x^+ -x^-)^2\Phi_h^2}\Big[\phi\Big(\frac{\partial^2 U}{\partial\Phi\partial\Psi^I}\Big\rvert_h\Big) + \psi^K\Big(\frac{\partial^2 U}{\partial\Psi^K\del\Psi^I}\Big\rvert_h\Big)\Big] = 0\ ,\quad \\
	&&\hspace{-4mm} \partial_+\partial_-\Omega + \frac{1}{(x^+ - x^-)^2}\Big[2\Omega - \frac{\phi}{\Phi_h}\Big(1+\frac{L^2}{4}\Big(\frac{\partial^2 U}{\partial\Phi\partial\Phi}\Big\rvert_h\Big)\Big) - \frac{L^2}{4\Phi_h}\Big(\frac{\partial^2 U}{\partial\Phi\partial\Psi^K}\Big\rvert_h\Big)\psi^K\Big] = 0\ . \nonumber
\end{eqnarray}
We define new scalar fields $\zeta^I = \psi^I - \beta^I\phi$, where $\beta^I$ are constants to be determined. Substituting in the linearized equations for $\psi^I$ above, we get decoupled equations for $\zeta^I$
\begin{equation}
(h_{IJ}|_h)\partial_+\partial_-\zeta^J+\frac{L^2}{(x^+ -x^-)^2\Phi_h^2}\Big(\frac{\partial^2 U}{\partial\Psi^K\del\Psi^I}\Big\rvert_h\Big)\zeta^K=0
\end{equation}
provided $\beta^I$ satisfy
\begin{equation}
	\Big[2(h_{IJ}|_h) - \frac{L^2}{\Phi_h^2}\Big(\frac{\partial^2 U}{\partial\Psi^I\partial\Psi^J}\Big\rvert_h\Big)\Big]\beta^J = \frac{L^2}{\Phi_h^2}\Big(\frac{\partial^2 U}{\partial\Phi\partial\Psi^I}\Big\rvert_h\Big)\ .
\end{equation}
Defining the matrix $H_{IJ} = 2(h_{IJ}|_h) - \frac{L^2}{\Phi_h^2}\Big(\frac{\partial^2 U}{\partial\Psi^I\partial\Psi^J}\Big\rvert_h\Big)$, we can solve for $\beta^J$ as
\begin{equation}
	\beta^I = H^{IJ}\Big[\frac{L^2}{\Phi_h^2}\Big(\frac{\partial^2 U}{\partial\Phi\partial\Psi^J}\Big\rvert_h\Big)\Big]\ ,
\end{equation}
where $H^{IJ}$ is inverse of $H_{IJ}$ (see Appendix \ref{z=1nonzerotheta}
for an example). With $h^{IJ}$ the inverse of $h_{IJ}$, the condition
for a stable $AdS_2$ critical point with no tachyonic or massless
modes is that the eigenvalues $m_I^2$ of the mass matrix\
$\frac{(h^{IJ}\vert_h)}{\Phi_h^2}(\frac{\partial^2 U}
{\partial\Psi^J\del\Psi^K}|_h)$\ satisfy the $AdS_2$
Breitenlohner-Freedman (BF) bound, \ie\ \ $m_I^2L^2\geq -{1\over 4}$\,.
Of course $m_I^2>0$ automatically satisfies this, as was the generic
case in \cite{Kolekar:2018sba}.\
For the case with simply one scalar field $\Psi$, the criteria for a
stable $AdS_2$ critical point are
\begin{equation}
U_h=0\ , \qquad \frac{\partial U}{\partial\Psi}\Big\rvert_h=0\ , \qquad \frac{\partial U}{\partial(\Phi^2)}\Big\rvert_h=\frac{-2}{L^2}\ , \qquad \frac{\partial^2 U}{\partial\Psi^2}\Big\rvert_h> -{\Phi_h^2\over 4L^2}\ .
\end{equation}

\section{Null energy conditions and a c-function}\label{neccfn}

We are studying 2-dim dilaton-gravity-matter theories (with a
potential) that we regard implicitly as arising from dimensional
reduction of higher dimensional gravity-matter theories. Requiring
time translations and that the space transverse to the two
$(t,r)$-dimensions is sufficiently symmetric means that the bulk space
effectively evolves nontrivially only in the bulk radial
direction. For instance, extremal branes enjoy translational and
rotational invariance in the spatial directions: these geometries thus
flow only in the radial direction.  From the dual point of view, with
the radial direction taken as encoding energy scales \cite{Susskind:1998dq,Peet:1998wn}, this simply
means that the theory has a nontrivial RG flow encoded by the bulk
theory in terms of a holographic renormalization group. This has been
the subject of much exploration with a large literature over the years
\eg\ \cite{Akhmedov:1998vf,Alvarez:1998wr,Girardello:1998pd,Distler:1998gb,
  Balasubramanian:1999jd,Skenderis:1999mm,deBoer:1999tgo,Verlinde:1999xm,
  deBoer:2000cz,Freedman:1999gp,Bousso:1999xy,Sahakian:1999bd,
  Goldstein:2005rr,Heemskerk:2010hk,Faulkner:2010jy}\ (and the recent
review \cite{Nishioka:2018khk}).

Focussing on reductions of extremal objects is equivalent to requiring
that the 2-dim theories approach an $AdS_2$ throat in the deep
infrared with the dilaton and scalars acquiring fixed point
values. The bulk radial flow to the infrared then must terminate at an
$AdS_2$ fixed point: the transverse space symmetries assumed above
imply that the bulk flow is effectively just 2-dimensional and the
dual theory is effectively encoded by a flow to a one dimensional
$CFT_1$ obtained by the dimensional reduction of the transverse
space. The bulk description of this holographic renormalization is
consistent with the reduction ansatz we have been discussing with the
size of the transverse space controlled by the 2-dimensional dilaton
$\Phi$.  It is important to note that this effective 2-dimensional
flow appears insensitive to whether the higher dimensional theory is
relativistic or nonrelativistic. In particular this raises the
question of proposing a c-theorem encoding the renormalization group
flow in the dual 1-dimensional theory. This is intriguing especially
considering that c-theorems and renormalization group flows are not so
easily constrained for nonrelativistic theories: if such a c-theorem
and associated c-function can be identified for the present context,
one may hope that the analysis here may aid progress in understanding
c-theorems for higher dimensional nonrelativistic theories away from
extremality. Previous investigations on holographic c-theorems in
Lifshitz and Schr\"{o}dinger theories can be found in
\cite{Liu:2012wf,Liu:2015xxa}.

From the bulk point of view, the gravitational theory is required
to satisfy appropriate energy conditions for being physically
well-defined. In particular the null energy conditions require
that the energy momentum tensor contracted with any null vector
$n^\mu$ be non-negative, \ie\ $T_{\mu\nu}n^\mu n^\nu\geq 0$. From the
Einstein equations governing the bulk theory (which is classical
in the large $N$ approximation), this imposes
$R_{\mu\nu}n^\mu n^\nu\geq 0$, which can be regarded as defining
monotonicity relations for bulk metric data. For relativistic
theories, there is a single null vector that is independent:
for nonrelativistic theories enjoying spatial translation symmetry,
there are two independent null vectors. The reduction ansatz we
have been discussing suggests a priori two independent null vectors,
one with components along the $(t,r)$ directions, the other with
components along $(t,x^i)$ directions.

Consider, for simplicity and concreteness, the ansatz for the
$D$-dim metric
\begin{equation}\label{metricansatz2}
  ds^2=-B^2 dt^2+\frac{du^2}{B^2}+\Phi^{\frac{4}{D-2}}\sum_{i=1}^{D-2}dx_i^2\ ,
\end{equation}
where $B$ and $\Phi$ depend only on the radial coordinate $u$ for the
sufficiently symmetric space we have in mind. We have chosen these
coordinates since the null energy conditions then simplify.
The components of the Ricci tensor are
\begin{eqnarray}\label{ricci2}
&& R_{tt}=B^2\Big[\frac{(B^2)''}{2}+\frac{2BB'\Phi'}{\Phi} \Big] \ , \qquad R_{uu}=-\frac{(B^2)''}{2B^2}-\frac{2}{\Phi B}(\Phi'B'+\Phi''B)+\frac{2(D-4)}{(D-2)}\frac{(\Phi')^2}{\Phi^2} \ , \nonumber \\[2pt]
&& R_{xx}=-\frac{\Phi^{\frac{4}{D-2}}}{(D-2)}\Big(\frac{4BB'\Phi'}{\Phi}+\frac{B^2(\Phi^2)''}{\Phi^2}\Big)\ ,
\end{eqnarray}
where prime ( $'$ ) denotes derivative w.r.t. the radial coordinate
$u$. For the two null vectors,
\begin{equation}\label{nullvecs2}
  \zeta^{M}=(\sqrt{-g^{tt}},\sqrt{g^{uu}},0,0,\dots,0)\ , \qquad
  \xi^{M}=(\sqrt{-g^{tt}},0,\sqrt{g^{xx}},0,\dots,0)\ ,
\end{equation}
the null energy conditions give
\begin{eqnarray}\label{necconstraints2}
&& R_{MN}\zeta^{M}\zeta^{N}=-2B^2\Big[\frac{\Phi''}{\Phi}-\frac{(D-4)}{(D-2)}\frac{(\Phi')^2}{\Phi^2}\Big] \geq 0\ , \nonumber\\[2pt]
&& R_{MN}\xi^{M}\xi^{N}=\frac{B^2}{2}\Big[\frac{(B^2)''}{B^2}-\frac{2}{(D-2)}\frac{(\Phi^2)''}{\Phi^2} +\frac{2(D-4)}{(D-2)}\frac{(B^2)'\Phi'}{B^2\Phi}\Big] \geq 0\ .
\end{eqnarray}
Note that the first condition is independent of $B$ in the coordinate
choice (\ref{metricansatz2}).

\noindent \emph{Example:}\ The charged finite temperature $D$-dim hvLif
metric is
\begin{equation}\label{hvLifmetric}
ds^2=\Big(\frac{r}{r_{hv}}\Big)^{-\frac{2\theta}{d_i}}\Big[-\frac{r^{2z}}{R^{2z}}
f(r) dt^2+\frac{R^2}{r^2 f(r)}dr^2+\frac{r^2}{R^2}\sum_{i=1}^{d_i}dx_i^2\Big]\ ,
\qquad d_i=D-2\ .
\end{equation}
The uncharged zero temperature case ($f(r)=1$) written in the form
\eqref{metricansatz2} has
\begin{eqnarray}\label{hvLifBPhi}
B^2(u)=\left(\frac{(z-\frac{2\theta}{d_i})^{2z-\frac{2\theta}{d_i}}R^{\frac{2\theta}{d_i}(z+1)-2z}}{r_{hv}^{\frac{2z\theta}{d_i}}}\right)^{\frac{1}{z-\frac{2\theta}{d_i}}}u^{\frac{2z-\frac{2\theta}{d_i}}{z-\frac{2\theta}{d_i}}}\ , \qquad &&
  u=\frac{r_{hv}^{2\theta/d_i}}{(z-2\theta/d_i)R^{z-1}}\,r^{z-\frac{2\theta}{d_i}}\ ,
  \nonumber \\
\Phi^2(u)=\left(\frac{(z-\frac{2\theta}{d_i})^{d_i-\theta}R^{3\theta-z\theta-d_i}}{r_{hv}^{2\theta-z\theta}}\right)^{\frac{1}{z-\frac{2\theta}{d_i}}}u^{\frac{d_i-\theta}{z-\frac{2\theta}{d_i}}}\ . \qquad && 
\end{eqnarray}
Substituting these expressions for $B^2$ and $\Phi^2$ in
\eqref{necconstraints2} recovers the familiar null energy conditions
for uncharged zero temperature hvLif theories
\begin{equation}\label{hvLifnec}
(d_i-\theta)(d_i(z-1)-\theta)\geq 0\ , \qquad (z-1)(d_i+z-\theta)\geq 0\ .
\end{equation}

\subsection{A holographic c-function}

The existence of a renormalization group flow in the radial direction
in the effective 2-dim bulk theory suggests the existence of a
c-function that encodes the number of degrees of freedom along the
flow. Requiring that the flow terminates at an $AdS_2$ fixed point
implies that the IR fixed point is a nontrivial $CFT_1$. The fact that
the $AdS_2$ is the very near horizon geometry of the extremal black
brane that describes the system suggests that the number of degrees of
freedom describing the IR $CFT_1$ is equal to the entropy of the
extremal black brane. The extremal entropy is given by the horizon
area
\be\label{extrEntropy}
S_{BH} = {g_{xx}^{(D-2)/2}|_h\, V_{D-2}\over 4G_D} = {\Phi_h^2\over 4G_2}\ ,
   \qquad\qquad G_2={G_D\over V_{D-2}}\ ,
\ee
with $\Phi_h$ the value of the dilaton (\ref{DdimKKmetric}) in the
$AdS_2$ region and $G_2$ the 2-dim Newton constant. Note that the dilaton
$\Phi$ controls the transverse area of the black brane.

This suggests proposing a holographic c-function after reduction 
of (\ref{metricansatz2}),
\be\label{cfn}
{\cal C}(u) = {\Phi^2(u)\over 4G_2} = {\Phi^2(u)\, V_{D-2}\over 4G_D}\ ,
\qquad\qquad \Phi^2=g_{xx}^{(D-2)/2}\ ,
\ee
describing the number of active degrees of freedom at scale $u$ along
the renormalization group flow to the IR $AdS_2$ fixed point. This was
proposed and discussed in \cite{Goldstein:2005rr} in the context of
4-dim nonsupersymmetric black hole attractors: in the present case, our
context is different in part but there is overlap in the physics nonetheless.

\begin{figure}[h] 
\hspace{0.5pc}
\includegraphics[width=16pc]{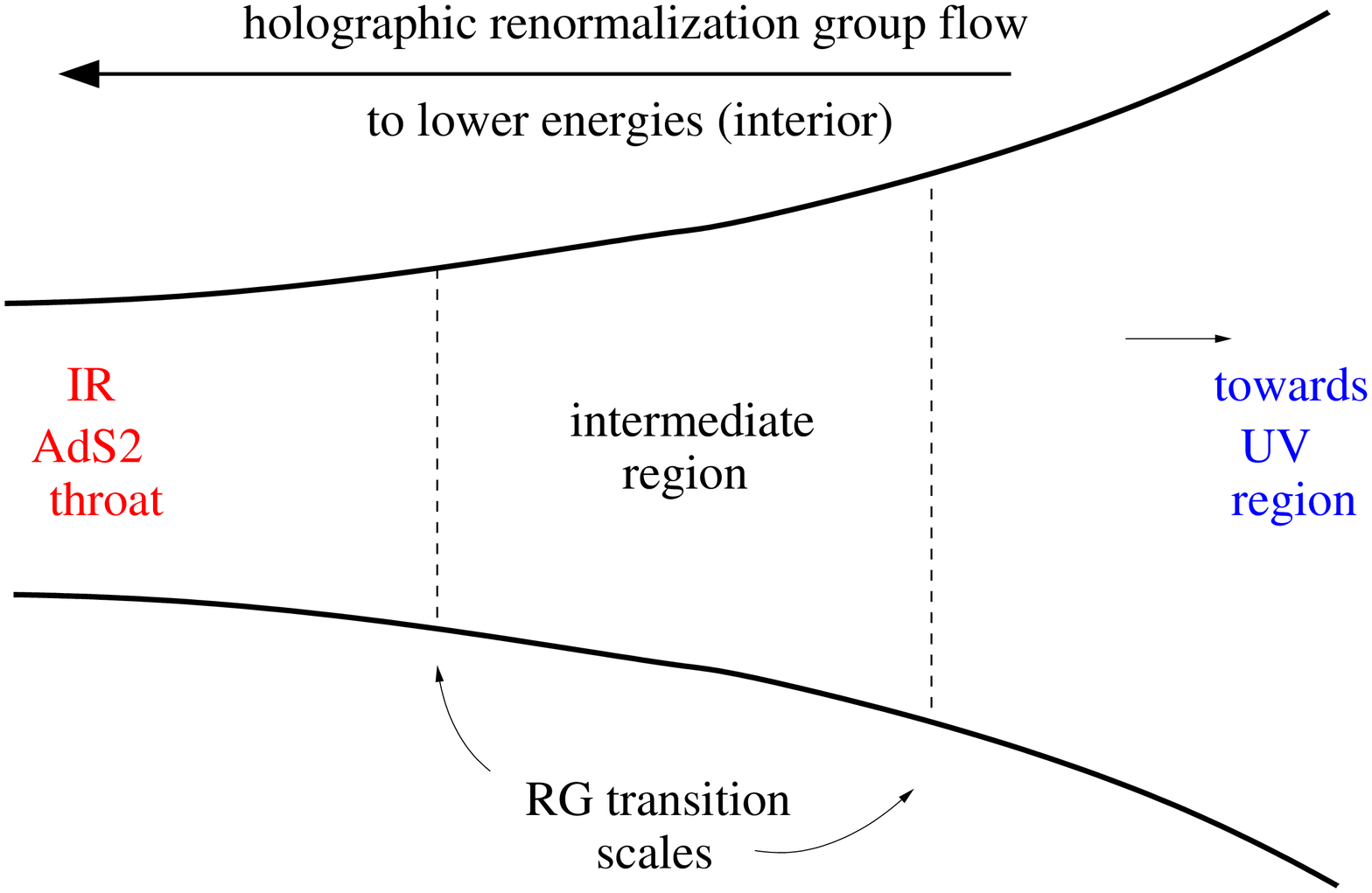} \hspace{2pc}
\begin{minipage}[b]{30pc}
  \caption{{\label{fig1}\footnotesize {A cartoon of the bulk spacetime
        with \newline the holographic RG flow in the radial direction to
        \newline the infrared $AdS_2$ throat region from the far (UV)
        \newline region through possible intermediate regions (and 
        \newline associated RG transition scales). \newline
}}}
\end{minipage}
\end{figure}
To prove the c-theorem for ${\cal C}$, we need to prove that ${\cal C}(u)$
decreases monotonically as we flow to lower energies $u$ (\ie\ interior),
or equivalently that ${\cal C}(u)$ is a monotonically increasing function
as $u$ increases (outwards to the boundary). We will do this in
two steps.

\noindent \underline{Step 1}:\ \
first define ${\tilde\Phi} = \Phi^{2/(D-2)}$.\ Then the first
of the energy conditions (\ref{necconstraints2}) becomes
\be\label{nec3} 
{\tilde\Phi} = \Phi^{{2\over D-2}}\ ;\quad
{\tilde\Phi}'' = {2\over D-2}\left({\Phi''\over\Phi} -
{D-4\over D-2} {(\Phi')^2\over\Phi^2}\right)\,\Phi^{{2\over D-2}}\quad
\Rightarrow\quad
-(D-2)\, B^2\, {{\tilde\Phi}''\over {\tilde\Phi}} \geq 0\ ,
\ee
so that ${\tilde\Phi}'$ monotonically decreases as $u$ increases
towards the boundary. In other words, ${\tilde\Phi}'$ in the interior
is larger than ${\tilde\Phi}'$ near the boundary. If we can now argue
that ${\tilde\Phi}'$ is positive near the boundary, this would imply
that ${\tilde\Phi}' > 0$ everywhere in the bulk as well. This then
would imply that ${\tilde\Phi}(u)$ is a monotonically increasing
function as $u$ increases and flows towards the boundary. A heuristic
picture of the setup appears in Figure~\ref{fig1}\ (see also
the discussion on nonconformal D2-branes, sec.~\ref{cfnD2M2}, which
exemplifies this).

\noindent \underline{Step 2}:\ \ Now we proceed to argue that
${\tilde\Phi}'$ is positive near the boundary for suitable boundary
conditions, namely that the ultraviolet of the theory belongs in the
hvLif family (\ref{hvLifmetric}) that we have been focussing on (which
also includes $AdS$ for exponents $z=1, \theta=0$). The extremal
branes we are considering here are excited states at finite charge
density in these theories: the near boundary region corresponds to the
high energy regime of the dual, well above the characteristic scales
of the excited states. So it suffices to use the asymptotic (uncharged
zero temperature) form of these backgrounds.

Using (\ref{hvLifBPhi}), we have $d_i=D-2$ and\
${\tilde\Phi}=\Phi^{\frac{2}{d_i}}$. Then retaining only relevant factors,
we have
\bea
&& \tilde{\Phi}\sim \Big(z-\frac{2\theta}{d_i}\Big)^{\frac{d_i-\theta}{zd_i-2\theta}}u^{\frac{d_i-\theta}{zd_i-2\theta}}\ , \qquad
\tilde{\Phi}'\sim \Big(z-\frac{2\theta}{d_i}\Big)^{\frac{d_i-\theta}{zd_i-2\theta}}\frac{(d_i-\theta)}{(zd_i-2\theta)}\frac{1}{u^{\frac{d_i(z-1)-\theta}{zd_i-2\theta}}}\ ,
\nonumber \\
&& \tilde{\Phi}''\sim -\Big(z-\frac{2\theta}{d_i}\Big)^{\frac{d_i-\theta}{zd_i-2\theta}}\frac{(d_i-\theta)(d_i(z-1)-\theta)}{(zd_i-2\theta)^2}\frac{1}{u^{\frac{d_i(z-1)-\theta}{zd_i-2\theta}+1}}\ .
\eea
Then $\frac{\tilde{\Phi}''}{\tilde{\Phi}}\leq 0$ gives the null energy
condition $(d_i-\theta)(d_i(z-1)-\theta)\geq 0$. A reasonable dual field
theory requires positivity of specific heat if the theory is excited to
finite temperature. Since the entropy for these theories scales as\
$S\sim V_{d_i} T^{d_i-\theta\over z}$, the positivity of the corresponding
specific heat imposes\ ${d_i-\theta\over z}\geq 0$.\ This implies
$d_i-\theta\geq 0$ since $z\geq 1$. Alongwith the null energy condition,
this leads to\ $(d_i(z-1)-\theta)\geq 0$.
These two conditions together imply
\be\label{zdi-2theta}
zd_i-2\theta = (d_i(z-1)-\theta)+(d_i-\theta)\geq 0\ .
\ee
Then we see that $\tilde{\Phi}'$ is positive in this near boundary region.
Roughly, ${\tilde\Phi}\sim u^n$ and ${\tilde\Phi}'\geq 0$ and
${\tilde\Phi}'' \leq 0$ require $n\geq 0$ and $n(n-1)\leq 0$, \ie\
$0\leq n\leq 1$. We have argued that this is true if the null energy
conditions and positivity of specific heat are satisfied.

Thus finally, we have shown that for the ultraviolet data we are
considering, ${\tilde\Phi}(u)=\Phi^{2/(D-2)}$ is monotonically
decreasing as $u$ flows to the interior (lower energies). Since the
exponent ${2\over D-2}$ is positive, this implies that $\Phi(u)$
satisfies the same monotonicity property. This proves that the
holographic c-function (\ref{cfn}) we propose in fact satisfies the
c-theorem.

At the IR $AdS_2$ horizon, ${\cal C}$ in (\ref{cfn}) approaches the
extremal black hole entropy (\ref{extrEntropy}), which is the IR
number of degrees of freedom controlling the number of black hole
microstates, akin to a central charge for this subsector. In fact it
is this requirement that ${\cal C}\ra S_{BH}$ at the IR $AdS_2$ fixed
point which fixes the precise scaling of ${\cal C}$ in terms of
$\Phi$\ (else any positive power of $\Phi$ is monotonic, from the
above arguments).

It is interesting to note that we have mainly used the first null energy
condition in (\ref{necconstraints2}) in the above arguments. The second
null energy condition appears to be more a condition on the matter
configurations: for instance, the second condition for hvLif backgrounds
(\ref{hvLifmetric}) gives $z\geq 1,\ d_i+z-\theta\geq 0$ in
\ref{hvLifnec}), which follow from reality of the
fluxes supporting the background, and also follows from specific
heat positivity.  to illustrate the condition in more generality,
let us restrict to $D=4$ for simplicity: then the second condition in
(\ref{necconstraints2}) gives
\be
\frac{(\Phi^2)''}{\Phi^2}\ \leq\  \frac{(B^2)''}{B^2} \ ,
\ee
which says that the dilaton ``acceleration'' is not greater than that
of the 2-dim metric. As we approach the $AdS_2$ region, we have\
$B^2\sim (u-u_0)^2$ so this becomes\
$\frac{(\Phi^2)''}{\Phi^2} \lesssim {2\over (u-u_0)^2}$\
which is trivially satisfied as $u\ra u_0$ since the right hand side
grows large. Thus the near $AdS_2$ region does not provide any
additional constraint from this energy condition. However the near
boundary region gives nontrivial constraints on the exponents defining
the theory from this energy condition as we have seen. We will discuss
this further later.

One might be concerned that the null energy conditions (and the
Einstein equations) are second order equations while renormalization
group flow is first order. It is important to note in this regard that
the boundary conditions we have imposed is on the first derivative
${\tilde\Phi}'$, which then automatically implies monotonicity. This
physical boundary condition has effectively ruled out the other
(growing) mode which would likely be singular in the interior.

In explicit examples (\eg\ nonconformal branes redux, later), we can
check this dilatonic c-function in fact has the right
behaviour. Consider for instance an extremal brane in an hvLif theory
where $B^2, \Phi^2$ near the boundary have the form (\ref{hvLifBPhi})
while in the near $AdS_2$ region, $B^2\sim (u-u_0)^2$ and $\Phi\sim
u^A$ globally, with $A={d_i-\theta\over zd_i-2\theta}$\,. Then using the
arguments around (\ref{zdi-2theta}), we see that $A\geq 0$ so that
$\Phi^2(u)$ can be seen explicitly to monotonically decrease through
the bulk as $u$ decreases flowing towards $AdS_2$. We also see that
$A\leq 1$ so that $\Phi''\leq 0$ in accord with the first energy
condition in (\ref{necconstraints2}).  The second energy condition in
the near boundary region simply imposes the constraints on the
exponents that we have seen, which are required of the theory. In the
near $AdS_2$ region, $B^2\sim (u-u_0)^2$ and so as described above,
the second energy condition is satisfied.  This family includes $AdS$
where $z=1, \theta=0$ and $\Phi^2=u^2$.

From the point of dual 1-dim theories which flow to the $CFT_1$ dual
to the $AdS_2$ bulk theory, the arguments above suggest that ${\cal C}$
in (\ref{cfn}) is a candidate c-function. While spatial
coarse-graining does not make sense in $0+1$-dim (no space!), the
renormalization group defined in terms of integrating out high energy
modes does make sense, \ie\ as a flow to lower energies (IR). In the
present context, we have defined the holographic c-function ${\cal C}$
as essentially inherited from the higher dimensional theory that has
been compactified: it would be interesting to understand the
c-function from the dual 1-dim point of view.

\subsection{Null energy conditions from the $2$-dim perspective}

We have described the null energy conditions $T_{\mu\nu}n^\mu n^\nu\geq 0$
in the higher dimensional theory and recast them in terms of 2-dim
bulk variables $g^{(2)}_{\mu\nu},\ \Phi$. The two independent null vectors
give two independent null energy conditions (\ref{necconstraints2})
as we have seen. However it is interesting to note that only one of
the null vectors --- $\zeta^M=(\sqrt{-g^{tt}}, \sqrt{g^{uu}})$ --- has a
natural interpretation intrinsically in the 2-dimensional spacetime.
This leads to the first of the energy conditions. The second one
appears to have no intrinsic interpretation directly in 2-dimensions:
however we can reverse engineer this from the higher dimensional
theory and recast it in terms of the potential governing the dilaton
and other scalars in the context of the dilaton-gravity-scalar theory
(\ref{2dimgsaction}).

The $tr-$ and $ii$-components, for $i=1,\dots,D-2$, of Einstein equations for the
gravity-scalar action in $D$-dimensions (\ref{Ddimgsaction}) are
\begin{equation}\label{EinsEqnD-tr}
\mathcal{R}^{(D)}_{\mu\nu}-\frac{g^{(D)}_{\mu\nu}}{2}\mathcal{R}^{(D)}=\frac{h_{IJ}}{2}\Big(\partial_{\mu}\Psi^I\partial_{\nu}\Psi^J-\frac{g^{(D)}_{\mu\nu}}{2}\partial_M\Psi^I\partial^M\Psi^J\Big)-\frac{g^{(D)}_{\mu\nu}}{2}V\ ,
\end{equation}
\begin{equation}\label{EinsEqnD-xx}
\mathcal{R}^{(D)}_{ii}-\frac{g^{(D)}_{ii}}{2}\mathcal{R}^{(D)}=\frac{h_{IJ}}{2}\Big(\partial_i\Psi^I\partial_i\Psi^J-\frac{g^{(D)}_{ii}}{2}\partial_M\Psi^I\partial^M\Psi^J\Big)-\frac{g^{(D)}_{ii}}{2}V +\frac{\partial V}{\partial g^{(D)\,ii}} \ ,
\end{equation}
where we have taken the metric to be diagonal in the spatial components \ie\ $g^{(D)}_{ij}=0\ \forall\ i\neq j$ and the potential $V$ in \eqref{Ddimgsaction} to be dependent only on $g^{(D)}_{ii}$ components. These equations $G^{(D)}_{MN}=8\pi G_D T^{(D)}_{MN}$ give the stress
tensor components as
\begin{eqnarray}
&& 8\pi G_D T^{(D)}_{\mu\nu}=\frac{h_{IJ}}{2}\Big(\partial_{\mu}\Psi^I\partial_{\nu}\Psi^J-\frac{g^{(D)}_{\mu\nu}}{2}\partial_M\Psi^I\partial^M\Psi^J\Big)-\frac{g^{(D)}_{\mu\nu}}{2}V\ , \\
&& 8\pi G_D T^{(D)}_{xx}=\frac{h_{IJ}}{2}\Big(\partial_x\Psi^I\partial_x\Psi^J-\frac{g^{(D)}_{xx}}{2}\partial_M\Psi^I\partial^M\Psi^J\Big)-\frac{g^{(D)}_{xx}}{2}V+\frac{\partial\,V}{\partial g^{(D)\,xx}}\ .\qquad
\end{eqnarray}
After dimensional reduction, we obtain the 2-dim action
(\ref{2dimgsaction}) and the above equations become the 2-dim Einstein
equations and the dilaton equation in (\ref{2dimgseoms}). In particular
the higher dimensional $\mu\nu$-components give 2-dim Einstein equations
which we write in the form
\begin{eqnarray}\label{2dEeqns}
&& \frac{1}{\Phi^2}[g_{\mu\nu}\nabla^2\Phi^2-\nabla_{\mu}\nabla_{\nu}\Phi^2]=8\pi G_DT^{(D)}_{\mu\nu}\ , \nonumber \\
&& 8\pi G_D T^{(D)}_{\mu\nu}=\frac{h_{IJ}}{2}\Big(\partial_{\mu}\Psi^I\partial_{\nu}\Psi^J-\frac{g_{\mu\nu}}{2}\partial_M\Psi^I\partial^M\Psi^J\Big)-\frac{g_{\mu\nu}U}{2\Phi^2}\ .
\end{eqnarray}
The $xx$-component of the higher dimensional stress tensor can likewise
be expressed in terms of the 2-dim potential and its derivative as
\begin{equation}
8\pi G_D T^{(D)}_{xx}=-\frac{\Phi^{\frac{4}{(D-2)}}}{4}h_{IJ}\partial_M\Psi^I\partial^M\Psi^J -\frac{(D-2)}{2}\Phi^{\frac{2}{D-2}+2}\,\frac{\partial U}{\partial\Phi^2}\ .
\end{equation}
This has no obvious 2-dim origin intrinsically: the
null energy condition intrinsic to two dimensions involves only
$T_{\mu\nu}$ but not $T_{xx}$. Further the null vector $\xi^M$ in
(\ref{nullvecs2}) has no intrinsically 2-dimensional meaning. The
second null energy condition in (\ref{necconstraints2}) from the
higher dimensional theory can however be recast in 2-dimensional
language in terms of the stress tensor components above and it is then
interesting to ask what 2-dimensional constraints it leads to. This
second NEC $T^{(D)}_{MN}\xi^M\xi^N\geq 0$ for the null vector
\begin{equation}\label{NEC2in2d}
\xi^M=(\sqrt{-g^{(D)\,tt}},0,\sqrt{g^{(D)\,xx}},0,\ldots,0) =
(\sqrt{-g^{tt}}\Phi^{\frac{D-3}{D-2}},0,\Phi^{\frac{-2}{D-2}},0,\dots,0)
\end{equation}
becomes
\begin{equation}
8\pi G_D(T^{(D)}_{tt}(\xi^t)^2+T^{(D)}_{xx}(\xi^x)^2)=-\Phi^{\frac{2(D-3)}{(D-2)}}\frac{g^{tt}h_{IJ}}{2}\partial_t\Psi^I\partial_t\Psi^J +\frac{\Phi^{\frac{-2}{D-2}}}{2}\Big(U-(D-2)\Phi^2\frac{\partial U}{\partial\Phi^2}\Big)\geq 0\ .
\end{equation}
For static backgrounds as we have here,\ $\partial_t\Psi^I=0$: then
this second NEC becomes a nontrivial condition on the potential and
its derivative
\begin{equation}\label{NEC2-UdU}
U - (D-2)\,\Phi^2\,\frac{\partial U}{\partial\Phi^2}\ \geq\ 0\ .
\end{equation}
In 2-dim dilaton-gravity-matter theories that arise from some higher
dimensional reduction, this condition (\ref{NEC2-UdU}) is simply
recognized as the second NEC in (\ref{necconstraints2}). However if we
regard (\ref{2dimgsaction}) as an intrinsically 2-dim theory, then 
it appears reasonable to impose such a constraint on the dilaton-matter
potential.

To illustrate this, consider first a potential of a form we have seen
arising from reduction of 4-dim Einstein-Maxwell theory,
\be
U=-V_0\Phi+{V_2\over\Phi^3}\qquad\Rightarrow\qquad U-2\Phi^2{dU\over d\Phi^2}
= {4V_2\over\Phi^4}\geq 0\quad\Rightarrow\quad V_2\geq 0\ .
\ee
Of course this can be recognized as the condition $Q^2\geq 0$ in the
higher dimensional theory: from the 2-dim point of view, the condition
gives positivity constraints on the coefficients that appear in the
potential.

The first NEC in $D$-dimensions, $T^{(D)}_{MN}\zeta^M\zeta^N\geq 0$, or
$\mathcal{R}^{(D)}_{MN}\zeta^M\zeta^N\geq 0$ for the null vector
$\zeta^M=(\sqrt{-g^{(D)\,tt}},\sqrt{g^{(D)\,rr}},0,0,\dots,0) =
(\sqrt{-g^{tt}}\Phi^{\frac{D-3}{D-2}},\sqrt{g^{rr}}\Phi^{\frac{D-3}{D-2}},0,0,\dots,0)$\  becomes the 2-dim NEC\ 
$\mathcal{R}_{\mu\nu}\tilde{\zeta}^{\mu}\tilde{\zeta}^{\nu}\geq 0$\ for
the 2-dim null vector\ $\tilde{\zeta}^{\mu}=(\sqrt{-g^{tt}},\sqrt{g^{rr}})$.\
Using \eqref{2dEeqns}, this NEC gives
\begin{equation}
-\nabla_{\mu}\nabla_{\nu}\Phi^2\tilde{\zeta}^{\mu}\tilde{\zeta}^{\nu}=g^{tt}\nabla_t\nabla_t\Phi^2-g^{rr}\nabla_r\nabla_r\Phi^2\geq 0\ .
\end{equation}
For static backgrounds, this recovers the condition on the
``acceleration'' of the dilaton that we have studied earlier in the
context of the c-theorem.

\subsection{The c-function in the $M2$-$D2$ system}\label{cfnD2M2}

Nonconformal $Dp$-branes upon dimensional reduction on the transverse
sphere give rise to hvLif theories with $z=1$ and nonzero $\theta$
\cite{Dong:2012se}. In particular the $D2$-brane supergravity phase
upon $S^6$-reduction give rise to bulk 4-dim hvLif theories with
$z=1,\ \theta=-{1\over 3}$\,. These flow \cite{Itzhaki:1998dd} in the
infrared to M2-branes, which give rise to $AdS_4$ upon
$S^7$-reduction\ (with $z=1, \theta=0$).
These are all uncharged phases. Adding a $U(1)$ gauge field to this
system --- which can be taken as the dual to the $U(1)_R$ current ---
and tuning to extremality gives string realizations for the extremal
versions of the above 4-dim theories. We have in mind that the
$AdS_4$ phase eventually terminates in the deep IR at an $AdS_2$ throat:
see Figure~\ref{fig1}.
Since the transition from the D2-phase to the M2-$AdS_4$ phase occurs
at energies well above the IR scale where the $AdS_2$ emerges, the
D2-phase can be essentially regarded as uncharged for the purposes of
the discussion below. In the far UV, the D2-branes are described
by free 3-dim SYM.

The string frame metric and the dilaton describing $N$ $D2$-branes are
\begin{eqnarray}
  ds_{st}^2 = \frac{r^{5/2}}{R_2^{5/2}}dx_{||}^2 + \frac{R_2^{5/2}}{r^{5/2}}dr^2
  + \frac{R_2^{5/2}}{r^{1/2}}d\Omega_6^2\ , \qquad
  e^{\phi} = g_s\Big(\frac{R_2^5}{r^5}\Big)^{1/4}\ ,
\end{eqnarray}
with $e^{\phi_{\infty}}=g_s$ the asymptotic value of the dilaton, and
\begin{equation}
g_{YM}^2=\frac{g_s}{\sqrt{\alpha'}}\ , \qquad R_2^5=\alpha'^3 g_{YM}^2 N\ ,
\end{equation}  
where we are ignoring numerical factors (since we will be primarily
interested here in the scaling behaviour along the RG flow).
The 10-dim Einstein frame metric
$ds_E^2=e^{-\frac{1}{2}(\phi-\phi_{\infty})}ds_{st}^2$ after dimensional
reduction on $S^6$ gives the Einstein metric of the effective 4-dim
hvLif theory with $d_i=2$, $z=1$, $\theta=-\frac{1}{3}$,
\begin{eqnarray}
ds^2 = \frac{r^{7/2}}{R_2^{7/2}}(-dt^2+dx_1^2+dx_2^2)+\frac{R_2^{3/2}}{r^{3/2}}dr^2
= \Big(\frac{\rho}{R_2}\Big)^{1/3}\Big[\frac{\rho^2}{R_2^2}(-dt^2+dx_i^2)
  + \frac{R_2^2}{\rho^2}d\rho^2\Big]\ .
\end{eqnarray}
The second expression is written in coordinates where the hvLif form
\eqref{hvLifmetric} is manifest. We have
\begin{equation}
  \rho = \frac{r^{3/2}}{R_2^{1/2}}\ ,\qquad w=\frac{r^{3/2}}{R_2^{5/2}}\ ,
  \qquad u={r^2\over R_2}\ ,
\end{equation}
where $w={r^{(5-p)/2}\over R_p^{(7-p)/2}}$ is the nonconformal $Dp$-brane
supergravity radius/energy
variable introduced in \cite{Peet:1998wn}. This coordinate has also
proved useful in studies of entanglement entropy and its interpretation 
in the nonconformal brane system \cite{Barbon:2008ut, Narayan:2013qga}.
The coordinate $u$ is chosen to cast the metric above in the form
(\ref{metricansatz2}) that we found useful in analysing the c-function
in our earlier discussion: in terms of those expressions, we have
\be
B^2 = {r^{7/2}\over R_2^{7/2}} = {u^{7/4}\over R_2^{7/4}}\ ,\qquad 
\Phi^2 = {r^{7/2}\over R_2^{7/2}} = w^{7/3}\, R_2^{7/3}\ .
\ee
Then the c-function (\ref{cfn}) written in terms of the energy variable
$w$ in the D2-phase is
\begin{equation}\label{D2cfn}
  {\cal C}(w) \sim {V_2 \Phi^2\over G_4} = V_2\, w^{7/3}\,
  N^2 {1\over(g_{YM}^2N)^{1/3}}\ =\ V_2\, w^2\, N_{eff}(w)\ ,
\end{equation}
after spatial compactification of the D2-branes. Here we have used
\begin{equation}
  G_4\sim \frac{G_{10}}{Vol(S^6)}\sim \frac{g_s^2\alpha'^4}{R_2^6}\ , \qquad
  N_{eff}(w)= N^2 {1\over(g_{YM}^2N/w)^{1/3}}\ ,
\end{equation}
and $N_{eff}(w)$ is the scale-dependent number of degrees of freedom
for the D2-phase (which also has played useful roles in entanglement
studies \cite{Barbon:2008ut, Narayan:2013qga}), while the dimensionless
gauge coupling at scale $w$ is $g_{eff}={g_{YM}^2N\over w^{3-p}}$\,.

The M2-phase is given by the $AdS_4\times S^7$ background (again ignoring
numerical factors)
\be
ds^2 = {r^2\over R^2} dx_{||}^2 + {R^2\over r^2} dr^2 + R^2 d\Omega_7^2\ ,
\qquad R^6\sim Nl_p^6\ ,\qquad
G_4\sim \frac{G_{11}}{Vol(S^7)}\sim \frac{l_p^9}{R^7}\ ,
\ee
which after reducing on the $S^7$ gives $AdS_4$\ (and $l_p$ is the 11-dim
Planck length; note that $R^6\sim Nl_P^6\sim g_sR_2^5\sqrt{\al'}$). This
is already in the form (\ref{metricansatz2}) with $u=r$ and the energy
variable\ $w={r\over R^2}$\ and\ $\Phi^2={r^2\over R^2}$\,.
Then the c-function in this M2-phase after spatial compactification is
\be\label{cfnM2}
{\cal C}(w) = {V_2\over G_4}\, {r^2\over R^2} = {R^2\over G_4}\, V_2\, w^2
   = N^{3/2}\, V_2\, w^2\ ,
\ee
using ${R^2\over G_4} = {R^9\over G_{11}} = N^{3/2}$.\ It is useful to
recall \cite{Itzhaki:1998dd} that the D2-phase is valid in the regime
$g_{YM}^2N^{1/5} \ll {r\over \al'} \ll g_{YM}^2N$\ so that\
$N^{3/2}\ll N_{eff}(w)\ll N^2$. At the scale
${r\over\al'}\sim g_{YM}^2$ the system transits from a smeared (arrayed)
M2-phase to the M2-$AdS_4$ phase. At this scale which corresponds to\
$w\sim g_{YM}^2 N^{-1/2}$, we have\
$N_{eff}(w) \sim N^{3/2}$\, and the D2-phase c-function can be seen to
match that in the M2-phase. The present analysis cannot be applied to
the intermediate interpolating phase corresponding to smeared (arrayed)
M2-branes.

We have so far discussed uncharged D2-M2 phases. With the $AdS_2$
emerging in the deep IR (within the $AdS_4$ region), the transition
between the D2- and M2-phase is well approximated by the uncharged
system. To see this explicitly, note that the charged hvLif metric
arising from D2-redux is
\begin{eqnarray}
ds^2&=&\Big(\frac{\rho}{R_2}\Big)^{1/3}\Big[\frac{\rho^2}{R_2^2}(-f(\rho)dt^2+dx_1^2+dx_2^2)+\frac{R_2^2}{\rho^2f(\rho)}d\rho^2\Big]\ , \nonumber \\
	f(\rho)&=& 1-\Big(\frac{\rho_0}{\rho}\Big)^{10/3}+\frac{Q_D^2}{\rho^{14/3}}\Big(1-\Big(\frac{\rho}{\rho_0}\Big)^{4/3}\Big)\ .	
\end{eqnarray}
with\ $Q_D^2\sim \rho_0^{7/3}$\ at extremality. Since the transition is
occurring at a scale $\rho_{trans}\gg \rho_0$, we essentially have
$f(\rho)\sim 1$ in that region.
In the deep IR, the extremal M2-$AdS_4$ phase
\begin{eqnarray}
ds^2 = {r^2\over R^2}(-f(r)dt^2+dx_i^2) + {R^2\over r^2 f(r)} dr^2\ ,\qquad
f(r) = 1-\Big(\frac{r_0}{r}\Big)^3+\frac{Q^2}{r^4}\Big(1-\frac{r}{r_0}\Big)\ ,
\end{eqnarray}
(after $S^7$ reduction) with $Q^2\sim r_0^4$ develops an $AdS_2$ throat,
with the horizon at $r=r_0$. Then the IR scale at the horizon is
$u={r_0\over R^2}$ and the c-function approaches 
\be
   {\cal C}\ \xrightarrow{\ AdS_2\ }\  N^{3/2}\, {V_2\, Q\over R^4}\ .
\ee
This phase is dual to a doped $CFT_3$, with dopant density
$\sigma_Q\equiv {Q\over R^4}$\ which is essentially the number of
dopant charge carriers per unit area of the M2-branes: then
${\cal C}_{IR}$ is essentially a ``central charge'' whose $N^{3/2}$
scaling reflects the underlying number of degrees of freedom of the
M2-$CFT$, which has been doped with an additional $V_2 \sigma_Q$
number of charge carriers distributed over the volume $V_2$ of the
M2-branes\ (there are some parallels with the heuristic partonic
picture of entanglement for excited $AdS$ plane wave states in
\cite{Mukherjee:2014gia}). This ``central charge'' corresponds to the
number of microstates of the doped $CFT_1$ obtained by spatial
compactification of the M2-branes: it is essentially dual to the
$AdS_2$ theory describing the extremal black brane with $C_{IR}$ the
extremal entropy.  In some sense, $w^2={Q(w)\over R^4}$ is a
scale-dependent dopant density with $w_{IR}^2={Q\over R^4}$ the
infrared value.  String/M-theory realizations of this involve turning
on an appropriate $G_4$-flux in the M2-brane system which after the
$S^7$ reduction gives the additional $U(1)$ gauge field that provides
charge \cite{Gauntlett:2007ma}.

It is clear that the c-function (\ref{D2cfn}) in the D2-phase gives a
larger number of degrees of freedom than that in the M2-phase
(\ref{cfnM2}) (noting the regimes for $w$, which flows to
lower energies), in accord with our general discussions of the
c-function earlier. This dovetails with the fact that $\theta$ is
negative in the D2-phase (with $z=1$). It is also worth noting that
the precise $N$-scalings etc arise from the precise dimensionful
factors contained in the dilaton.

\subsection{c-function in the $M5$-$D4$ system}\label{cfnM5D4}

The M5-D4 brane system flows from an $AdS_7\times S^4$ phase (dual to
the 6-dim $(2,0)$ theory) through the D4-supergravity phase to finally
5-dim SYM in the IR. While this does not admit an $AdS_2$ region in
the deep IR of the phase diagram, it is interesting to study the
c-function (\ref{cfn}) in this case as well. This discussion has 
parallels with the D2-M2 case so we will be succinct.\
For the $M5$-$AdS_7$ phase, we have
\be
ds^2 = {r^2\over R^2} dx_{||}^2 + {R^2\over r^2} dr^2 + R^2 d\Omega_4^2\ ,
\qquad R^3\sim Nl_p^3\ ,\qquad
G_7\sim \frac{G_{11}}{Vol(S^4)}\sim \frac{l_p^9}{R^4}\ ,
\ee
which after reducing on the $S^4$ gives $AdS_7$\ ($l_p$ is the 11-dim
Planck length). This dovetails with (\ref{metricansatz2}) with $u=r$
and the energy variable\ $w={r\over R^2}$\ and\ $\Phi^2={r^2\over R^2}$\,.
The c-function in this M5-phase after spatial $T^5$-compactification then is
\be\label{cfnM5}
{\cal C}(w)_{M5} = {V_5\over G_7}\, {r^2\over R^2}
     = {R^5\over G_7}\, V_5\, w^5 = N^3\, V_5\, w^5\ ,
\ee
using ${R^5\over G_7} \sim {R^9\over G_{11}} \sim N^3$ here.\
Now for the $N$ $D4$-branes phase, the string metric and dilaton are
\begin{eqnarray}
ds_{st}^2 = \frac{r^{3/2}}{R_4^{3/2}}dx_{||}^2 + \frac{R_2^{3/2}}{r^{3/2}}dr^2
  + R_2^{3/2}\,r^{1/2}d\Omega_4^2\ , &&
  e^{\phi} = g_s\Big(\frac{r}{R_4}\Big)^{3/4}\ ,\nonumber\\
   g_{YM}^2\sim g_s\sqrt{\alpha'} , &&\ R_4^3\sim\alpha' g_{YM}^2 N\ ,
\end{eqnarray}
ignoring numerical factors. The 10-dim Einstein frame metric
$ds_E^2=e^{-\frac{1}{2}(\phi-\phi_{\infty})}ds_{st}^2$ after $S^4$-redux gives
a 6-dim hvLif theory \eqref{hvLifmetric} with $d_i=4$, $z=1$, $\theta=-1$,
\begin{eqnarray}
ds^2 = \frac{r^{5/4}}{R_4^{5/4}}\Big(-dt^2+\sum_{i=1}^4 dx_i^2\Big)+\frac{R_4^{7/4}}{r^{7/4}}dr^2
= \Big(\frac{\rho}{R_4}\Big)^{1/2}\Big[\frac{\rho^2}{R_4^2}(-dt^2
  +\sum_{i=1}^4dx_i^2) + \frac{R_4^2}{\rho^2}d\rho^2\Big] .\ \
\end{eqnarray}
The D4-brane supergravity radius/energy variable $w$ \cite{Peet:1998wn} and
the $u$ coordinate in (\ref{metricansatz2}) are
\be
\rho = R_4^{1/2} r^{1/2}\ ,\qquad w=\frac{r^{1/2}}{R_4^{3/2}}\ ,\qquad
u = {r^{3/2}\over R_4^{1/2}}\ .
\ee
With $G_6\sim G_{10}/Vol(S^4)\sim g_s^2\al'^4/R_4^4$,\  the
scale-dependent number of degrees of freedom $N_{eff}(w)$ for the
D4-phase \cite{Barbon:2008ut, Narayan:2013qga}, and the dilaton\
$\Phi^2=g_{xx}^{(D-2)/2}={r^{5/2}\over R_4^{5/2}}$, the c-function (\ref{cfn})
is
\be
   {\cal C}(w)_{D4} = {V_4\Phi^2(w)\over 4G_6} \sim V_4\, w^4\, N_{eff}(w)\ ,
   \qquad N_{eff}(w) = N^2 (g_{YM}^2N w)\ ,
\ee
after spatial $T^4$-compactification, and the regime of validity is\
$1\ll g_{YM}^2N w \ll N^{2/3}$\,.\ Noting\
$R_{11}=g_s\sqrt{\al'}\sim g_{YM}^2$ and $V_5=V_4R_{11}$, we see that
the c-function continuously transits from the M5- to the D4-phase for
length scales longer than the 11th circle size $R_{11}$. This leads to
the guess that the c-function in the free 5-dim SYM phase after
spatial $T^4$-compactification is possibly\ $N^2V_4w^4$.

\subsection{On dilatonic and entropic c-functions}\label{Entrcfn}

It is interesting to compare the dilatonic c-function we have defined
with the entropic c-function \cite{Casini:2006es,Casini:2012ei} that
has been studied based on studies of entanglement entropy
\cite{Myers:2010xs,Myers:2010tj,Myers:2012ed,Cremonini:2013ipa}.

Consider the bulk geometry (\ref{metricansatz2}) with asymptotics
being $AdS$ or hvLif, focussing on $D=4$ dimensions (with no
compactification). For a strip subsystem with width along say $x$,
the induced metric on a time slice is $\Phi^2dy^2+{du^2\over B^2}+\Phi^2dx^2$
and the area functional for holographic entanglement \cite{Ryu:2006bv} is\
$A = L\int du\, {\Phi\over B}\sqrt{1+B^2\Phi^2\, ({dx\over du})^2}$\
which after extremization gives
\be
S = {2L\over 4G_4}\int {du\over B} {\Phi^3\over\sqrt{\Phi^4-\Phi_*^4}}\ ,
\qquad l = \int {du\over B} {\Phi_*^2\over\Phi\sqrt{\Phi^4-\Phi_*^4}}\ ,
\ee
where $\Phi_*=\Phi(u_*)$ is the value of the dilaton at the turning
point $u_*$ of the minimal surface, and $L$ is the size of the
(essentially infinitely long) strip in the longitudinal $y$ direction.
For instance, for a strip in $AdS_4$, the area and width integrals
are\ $S={2L\over 4G_4}\int_\epsilon^{u_*} {Rdu\over u}
{u^3\over\sqrt{u^4-u_*^4}}\sim {R^4\over G_4} ({L\over\epsilon}-{L\over l})$\
and $l\sim {R^2\over u_*}$\,. The entropic c-function is then defined as
\be\label{entrcfn}
c_E = {l^2\over L} {dS\over dl}\ ,
\ee
which gives\ $c_E \sim {R^2\over G_4}$\,.
This is thus a measure of the local number of degrees of freedom, or
central charge, in the dual field theory responsible for
entanglement. In theories with an RG flow, the entropic c-function is
scale dependent and satisfies\ ${dc_E\over dl}\leq 0$, \ie\ it
monotonically decreases with the width $l$, and thus plays the role of
a c-function based on entanglement entropy. For instance for
nonconformal branes, $c_E(l)\sim N_{eff}(l)$.

We will now try to draw comparisons between this entropic c-function
and the dilatonic c-function (\ref{cfn}). Away from the $AdS_2$
horizon, $u\gg u_0$ and we have\ $B\sim {u-u_0\over R}\sim {u\over R}$\,.
Since the dilaton monotonically decreases flowing towards the interior,
\ie\ as $u$ decreases in (\ref{metricansatz2}), we can recast the
integrals above as
\be
S = {2L\over 4G_4}\, \Phi_*\, R \int_{\varphi_\epsilon}^1 {du\over u}
{\varphi^3\over\sqrt{\varphi^4-1}}\ ,\qquad
l = {2R\over\Phi_*} \int_{\varphi_\epsilon}^1 {du\over u}
{1\over\varphi\sqrt{\varphi^4-1}}\ ,\qquad \varphi={\Phi\over\Phi_*}\ .
\ee
Since the dilaton decreases monotonically, we can redefine the radial
variable by $\varphi$ as $du={d\varphi\over\varphi'}$\,, and we note
that all the information about the turning point has been scaled out
after this recasting to the factors outside the integrals.
The entropic c-function receives a nonvanishing contribution simply
from the finite part for which the integrals are simply finite
numerical factors. Then we see that\ $S\sim {R\over G_4} \Phi_*L\sim
{R^2\over G_4}\,{L\over l}$ which recovers $c_E\sim {R^2\over G_4}$\,.

From the discussion of the c-function for M2-branes (\ref{cfnM2}), we
see that ${\cal C}(w) = N^{3/2} V_2 w^2$. Recalling that ${R^2\over
  G_4}\sim N^{3/2}$, we see that the dilatonic c-function scales as
the entropic number of degrees of freedom (\ie\ $c_E$), but is in
addition extensive: it scales with $V_2$ and shrinks as $w^2$ along
the flow to the IR. In the 2-dim bulk theory after compactification,
$c_E$ cannot be formulated since the spatial directions are
compactified but the dilatonic c-function nevertheless encodes the
number of degrees of freedom that $c_E$ encodes. Similar comparisons
can be drawn for other cases.

It is also interesting to recall the holographic c-function in 
\cite{Freedman:1999gp}. For a bulk theory\
$ds^2=e^{2A(\varrho)} dx_\parallel^2+d\varrho^2$
enjoying Lorentz invariance (\ie\ $z=1$), this c-function is\
$C_{FGPW}(w)\sim {1\over G_D\, (dA/d\varrho)^{d_i}}$ where $d_i$ is the
number of boundary spatial dimensions. For $AdS_D$, we have\
$A\sim \log{\varrho\over R}$ and\ $c_{FGPW}\sim {R^{d_i}\over G_D}$\
which gives\ $c_{FGPW}\sim N^{3/2}$ for M2-$AdS_4$.\
For nonconformal $Dp$-branes\ ($g_{YM}^2\sim g_s {\al'}^{(p-3)/2}$)\
\cite{Itzhaki:1998dd}
\be
ds^2_{st} = {r^{(7-p)/2}\over R_p^{(7-p)/2}} dx_{\parallel}^2 
+ {R_p^{(7-p)/2}\over r^{(7-p)/2}} (dr^2 + r^2d\Omega_{8-p}^2),\quad 
e^\Phi = g_s \Big({R_p^{7-p}\over r^{7-p}}\Big)^{{3-p\over 4}} ,
\ \ \ R_p^{7-p}\sim g_{YM}^2N {\al'}^{5-p} ,
\ee
upon $S^{8-p}$-redux give the hvLif metric (\ref{hvLifmetric}) with
$z=1,\ p=d_i$\ \cite{Dong:2012se}:\ in this case, using
\be
\theta=p-{9-p\over 5-p}=-{(p-3)^2\over 5-p}\ ,\quad
ds_{p+2}^2=\Big({\varrho\over R_p}\Big)^{2\left(1-{d_i\over\theta}\right)}
dx_\parallel^2+d\varrho^2\ ,\quad
{\varrho\over R_p} = \Big({r\over R_p}\Big)^{-{\theta (5-d_i)\over 2d_i}} ,
\ee
we see that the dilatonic c-function (\ref{cfn}) we have discussed
(after redux to 2-dim) gives
\be
{\cal C}(w) \sim N_{eff}(w) V_{d_i} w^{d_i}\ ,
\qquad 
N_{eff}(w) = N^2 \left({g_{YM}^2N\over w^{3-p}}\right)^{p-3\over 5-p} ,
\quad w={r^{(5-p)/2}\over R_p^{(7-p)/2}}\ ,
\ee
as we have seen earlier in the detailed discussions on the D2-M2 and
M5-D4 phases (with $w$ the nonconformal $Dp$-brane supergravity
radius/energy variable \cite{Peet:1998wn}). On the other hand, the
c-function in \cite{Freedman:1999gp} mentioned above can be seen to
be\ $c_{FGPW}\sim N_{eff}(w)$,\ with the same scaling as the entropic
c-function $c_E$: this is a measure of the local degrees of freedom
of the higher dimensional theory, while the dilatonic c-function
${\cal C}$ has additional extensivity arising from the compactification.

\section{2-dim radial Hamiltonian formalism and $\beta$-functions}\label{dVV2d}

A version of the holographic renormalization group was formulated in
\cite{deBoer:1999tgo}: using a radial ADM-type split of the bulk
spacetime, the radial Hamiltonian constraint gives rise to flow
equations for couplings and corresponding $\beta$-functions. This is
not a Wilsonian formulation since the effective action at the scale
corresponding to some radial slice depends on data not just at higher
energy scales that have been integrated out: Wilsonian formulations of
the holographic renormalization group have been investigated in
\cite{Heemskerk:2010hk,Faulkner:2010jy}.  Nevertheless this dVV
formulation gives useful qualitative insights into the holographic
renormalization group. In this section, we will adapt this to obtain
renormalization group flow equations and $\beta$-functions starting
with the 2-dim dilaton-gravity-scalar theory. As in
\cite{deBoer:1999tgo}, writing the boundary 1-dim action on some
radial slice in a radial Hamilton-Jacobi formulation, we separate this
at low scales into local and nonlocal parts and then write the local
part in a derivative expansion. Taking the leading term to arise from
just a ``boundary potential'' term for the couplings (scalars
$\Phi,\Psi^I$), \ie\ no derivatives, we obtain relations between the
original potential and the boundary potential using the Hamiltonian
constraint, thereby obtaining $\beta$-functions from the flow
equations. We will describe this below.

Consider the 2-dim gravity-scalar action \eqref{2dimgsaction}
including the Gibbons-Hawking term
\begin{equation}\label{2dimgsactionwithGH}
  S=\frac{1}{16\pi G_2}\int  d^2x\sqrt{-g}\Big(\Phi^2\mathcal{R}-\frac{\Phi^2}{2}h_{IJ}\partial_{\mu}\Psi^I\partial^{\mu}\Psi^J-U(\Phi,\Psi^I) \Big)+\frac{1}{16\pi G_2}\int dt\sqrt{-\gamma}\Phi^2 2K\ ,
\end{equation}
where $U(\Phi,\Psi^I)=V\Phi^{\frac{2}{D-2}}$\,. Substituting the radial
decomposition of the metric
\begin{equation}\label{2dmetricfoliation}
ds^2=(N^2+\gamma_{tt}(N^t)^2)dr^2+2\gamma_{tt}N^tdtdr+\gamma_{tt}dt^2\ ,
\end{equation}
certain boundary terms cancel with the Gibbons-Hawking term: then
massaging leads to a radial Lagrangian (in Appendix
\ref{2dradialLderivation}, we derive this from dimensional reduction
of the Hamiltonian formulation of a higher dimensional theory of the
sort we have been considering)
\begin{equation}\label{2dradialL}
L=\frac{1}{16\pi G_2}\int dt\sqrt{-\gamma}\,N\, \Big[-2\Box_t\Phi^2+\frac{2K}{N}(\partial_r\Phi^2-N^t\partial_t\Phi^2)-\frac{\Phi^2}{2}h_{IJ}\partial_{\mu}\Psi^I\partial^{\mu}\Psi^J-U\Big]\ ,
\end{equation}
where the extrinsic curvature and the covariant derivative w.r.t.
$\gamma_{tt}$ are
\begin{equation}
K_{tt}=\frac{1}{2N}(\partial_r\gamma_{tt}-2D_tN_t)\ , \quad K=\gamma^{tt}K_{tt}\ , \quad D_tN_t=\partial_tN_t-\Gamma^t_{tt}N_t\ , \quad \Box_t\equiv \gamma^{tt}D_t D_t\ ,
\end{equation}
where $\gamma^{tt}=(\gamma_{tt})^{-1}$ and $N_t=\gamma_{tt}N^t$. The conjugate
momenta for the fields $\gamma_{tt}$, $\Phi$ and $\Psi^I$ are
\begin{eqnarray}\label{2dmomenta}
	&& \pi^{tt}\equiv \frac{16\pi G_2}{\sqrt{-\gamma}}\frac{\delta L}{\delta\dot{\gamma}_{tt}}=\frac{\gamma^{tt}}{N}(\partial_r\Phi^2-N^t\partial_t\Phi^2)\ , \nonumber \\
	&& \pi_{\Phi}\equiv \frac{16\pi G_2}{\sqrt{-\gamma}}\frac{\delta L}{\delta\dot{\Phi}}=4K\Phi=\frac{2\Phi\gamma^{tt}}{N}(\dot{\gamma}_{tt}-2D_tN_t)\ , \nonumber \\
	&& \pi_{I}\equiv \frac{16\pi G_2}{\sqrt{-\gamma}}\frac{\delta L}{\delta\dot{\Psi}^I}=-\frac{\Phi^2 h_{IJ}}{N}(\dot{\Psi}^J-N^t\partial_t\Psi^J)\ ,
\end{eqnarray}
where dot represents the radial derivative,\ie\
$\dot{\Phi}=\partial_r\Phi$ and so on. Inverting, we obtain
\begin{eqnarray}\label{2dfloweqns}
	&& \dot{\Phi}=\frac{1}{2\Phi}\Big(\frac{N\pi^{tt}}{\gamma^{tt}}+N^t\partial_t\Phi^2\Big)\ , \nonumber \\
	&& \dot{\gamma}_{tt}=\frac{N\pi_{\Phi}}{2\Phi\gamma^{tt}}+2D_tN_t\ , \nonumber \\
	&& \dot{\Psi}^I=-\frac{Nh^{IJ}\pi_J}{\Phi^2}+N^t\partial_t\Psi^I\ .
\end{eqnarray}
The Hamiltonian is obtained by a Legendre transform of the Lagrangian
\eqref{2dradialL} as
\begin{equation}
	H=\frac{1}{16\pi G_2}\int dt\sqrt{-\gamma}(\pi^{tt}\dot{\gamma}_{tt}+\pi_{\Phi}\dot{\Phi}+\pi_{I}\dot{\Psi}^I)-L=\frac{1}{16\pi G_2}\int dt \sqrt{-\gamma}(N\mathcal{H}+N^t\mathcal{H}_t)\ .
\end{equation}
The fields $N$ and $N^{t}$ being non-dynamical gives the constraints $\frac{\partial H}{\partial N}=0$ and $\frac{\partial H}{\partial N^t}=0$ \ie\
\begin{eqnarray}
	&& \mathcal{H}=\frac{\pi^{tt}\pi_{\Phi}}{2\Phi\gamma^{tt}}+2\Box_t\Phi^2+U-\frac{\pi^I\pi_I}{2\Phi^2}+\frac{\Phi^2}{2}h_{IJ}\gamma^{tt}\partial_t\Psi^I\partial_t\Psi^J=0\ , \label{2dNconstraint}\\
	&& \mathcal{H}_t=-2\gamma_{tt}D_t\pi^{tt}+\pi_{\Phi}\partial_t\Phi+\pi_I\partial_t\Psi^I=0\ . \label{2dNtconstraint}
\end{eqnarray}
Now as in \cite{deBoer:1999tgo}, we imagine that the boundary action
on some radial slice can be evaluated as a function of boundary field
values at that scale: then thinking of this action in terms of a radial
Hamilton-Jacobi formulation allows us to relate the conjugate momenta
as derivatives of this action, which we then use in the Hamiltonian
constraints in a derivative expansion to relate the bulk and boundary
expressions. Towards this, we segregate this boundary action into
a local part and a nonlocal part at a low energy scale $\mu\ll\mu_c$
(with $\mu_c$ the UV cut-off) in a derivative expansion,
\begin{equation}
S_{bdy}=S_{loc}+\Gamma\ .
\end{equation}
Here $\Gamma$ contains higher derivative, nonlocal terms which encode
the information about correlation functions of the operators in the
boundary theory and gives flow equations for these correlation
functions, which are the Callan-Symanzik equations (we will not explore
that here). A general form of the local action is\
$S_{loc}=\int dt \sqrt{-\gamma}\, \left( W(\Phi,\Psi^I) + \frac{M_{IJ}}{2}\partial_a\Psi^I\partial^a\Psi^J   + \ldots \right)$,   
with $W$, $M_{IJ}$ are local functions of the couplings.
Approximating the local part of the boundary action in terms of just
the leading potential term (ignoring derivatives) as
\begin{equation}
S_{bdy}=\int dt\sqrt{-\gamma}\,W(\Phi,\Psi^I)+\cdots\ ,
\end{equation}
we can define the conjugate momenta in terms of the boundary potential
$W(\Phi,\Psi^I)$ as
\begin{eqnarray}\label{2dmomentabdy}
&& \pi^{tt}\equiv \frac{16\pi G_2}{\sqrt{-\gamma}}\frac{\delta S_{bdy}}{\delta\gamma_{tt}}=8\pi G_2\gamma^{tt}\,W \ , \nonumber \\
&& \pi_{\Phi}\equiv \frac{16\pi G_2}{\sqrt{-\gamma}}\frac{\delta S_{bdy}}{\delta\Phi}=16\pi G_2\frac{\partial W}{\partial\Phi}\ , \nonumber \\
&& \pi_{I}\equiv \frac{16\pi G_2}{\sqrt{-\gamma}}\frac{\delta S_{bdy}}{\delta\Psi^I}=16\pi G_2\frac{\partial W}{\partial\Psi^I}\ .
\end{eqnarray}
Substituting these momenta in the Hamiltonian constraint
\eqref{2dNconstraint} and collecting the potential terms, we get a
relation between the bulk potential $U$ and the boundary potential $W$ as
\begin{equation}\label{2dpotentialsrelation}
\frac{U}{(8\pi G_2)^2}=\frac{2h^{IJ}}{\Phi^2}\frac{\partial W}{\partial\Psi^I}\frac{\partial W}{\partial\Psi^J}-\frac{W}{\Phi}\frac{\partial W}{\partial\Phi}\ .
\end{equation}
Using the momenta \eqref{2dmomentabdy} in terms of $W$, the flow equations
\eqref{2dfloweqns} can be written as
\begin{eqnarray}\label{2dfloweqnsbdy}
&& \dot{\Phi}=\frac{1}{2\Phi}\Big((8\pi G_2) NW+N^t\partial_t\Phi^2\Big)\ , \nonumber \\
&& \dot{\gamma}_{tt}=\frac{(8\pi G_2)N}{\Phi\gamma^{tt}}\frac{\partial W}{\partial\Phi}+2D_tN_t\ , \nonumber \\
&& \dot{\Psi}^I=-\frac{(16\pi G_2)Nh^{IJ}}{\Phi^2}\frac{\partial W}{\partial\Psi^J}+N^t\partial_t\Psi^I\ .
\end{eqnarray}

\noindent\underline{$\beta$-functions:}\ \
Choosing Fefferman-Graham gauge
\begin{equation}
N=1\ , \qquad  N^t=0\ ; \qquad ds^2=dr^2+\gamma_{tt}dt^2\ ,
\end{equation}
the flow equations become
\begin{equation}\label{2dfloweqnsbdyFGgauge}
\dot{\Phi}=\frac{(4\pi G_2)\,W}{\Phi}\ , \qquad \dot{\gamma}_{tt}=(8\pi G_2)\frac{\gamma_{tt}}{\Phi}\frac{\partial W}{\partial\Phi}\ , \qquad \dot{\Psi}^I=-\frac{(16\pi G_2)h^{IJ}}{\Phi^2}\frac{\partial W}{\partial\Psi^J}\ .
\end{equation}
From the above equation, we see that we can split the radial and time
dependence of $\gamma_{tt}$ as $\gamma_{tt}=a^2\hat{\gamma}_{tt}$,
where $a=a(r)$ and $\hat{\gamma}_{tt}$ is independent of $r$ (\ie\
$\gamma_{tt}$ simply rescales under RG flow). Then the flow equation
for $\gamma_{tt}$ gives
\begin{equation}
\dot{a}=\frac{(4\pi G_2)}{\Phi}\frac{\partial W}{\partial\Phi}a\ .
\end{equation}
Using this relation, we can write the radial derivatives in terms of
$\dot{a}$.\ In contrast with the higher dimensional cases in
\cite{deBoer:1999tgo}, note that this brings a factor of
${\del W\over\del\Phi}$ in the $\beta$-functions, which we define
for the RG flow as
\begin{eqnarray}
&& \beta^I\equiv a\frac{d}{da}\Psi^I=\frac{a\dot{\Psi}}{\dot{a}}=-\frac{4h^{IJ}}{\Phi\, \frac{\partial W}{\partial \Phi}}\frac{\partial W}{\partial \Psi^J}\ , \label{2dbetapsi} \\
&& \beta_{\Phi}\equiv a\frac{d}{da}\Phi=\frac{a\dot{\Phi}}{\dot{a}}=\frac{W}{\frac{\partial W}{\partial \Phi}}\ . \label{2dbetaphi}
\end{eqnarray}
We can write the relation \eqref{2dpotentialsrelation} between $U$ and
$W$ in terms of $\beta$-functions as
\begin{equation}
\frac{U}{(8\pi G_2)^2}=\frac{W^2h_{IJ}\beta^I\beta^J}{8\beta_{\Phi}^2}-\frac{W^2}{\Phi\beta_{\Phi}}\ .
\end{equation}

\subsection{$\beta$ functions for conformal/non-conformal theories}

In this subsection, we have set all the scales to unity \ie\ $R=1$,
$r_{hv}=1$, $8\pi G_2=1$. 

\noindent\underline{$\beta$-functions, conformal branes}:\ \ 
The effective potential for $2$-dim dilaton-gravity theories obtained
from \eg\ reductions of conformal branes is of the form $U(\Phi)$:
then from \eqref{2dpotentialsrelation}, the boundary potential is given by
\begin{equation}\label{WPhiconf}
\frac{dW^2}{d\Phi^2}=-U \qquad \implies\qquad
W^2=-\int_{\Phi_h}^{\Phi} U\, d\Phi^2 \ ,
\end{equation}
where we have imposed $W^2(\Phi_h)=0$. This boundary condition in a
sense reflects the fact that the 1-dim background corresponds to
zero energy. Expanding $U$ around the critical point,
\begin{equation}
U = \Big(\frac{dU}{d\Phi^2}\Big\vert_h\Big) (\Phi^2-\Phi_h^2) + \cdots\ ,
\qquad \text{where}\quad U\vert_h=0\ ,
\end{equation}
the $\beta$-function becomes
\begin{equation}
\beta_{\Phi}=\frac{W^2}{\Phi\frac{d W^2}{d\Phi^2}}=\frac{\int_{\Phi_h}^{\Phi} d\Phi^2 U}{\Phi U}=\frac{\Big(\frac{\Phi^4}{2}-\Phi^2\Phi_h^2\Big)+\Big(\frac{\Phi_h^4}{2}\Big)}{\Phi_h(\Phi^2-\Phi_h^2)}=\frac{(\Phi^2-\Phi_h^2)^2}{2\Phi_h(\Phi^2-\Phi_h^2)}=\frac{(\Phi^2-\Phi_h^2)}{2\Phi_h}\ .
\end{equation}
At the critical point, $\Phi=\Phi_h$, we see that $\beta_{\Phi}$
vanishes, consistent with the expectation that the $AdS_2$ critical
point background arises at the fixed point of the RG flow.

\noindent\underline{$\beta$-functions, nonconformal branes}:\ \
The effective potential for $2$-dim dilaton-gravity-scalar theories
obtained from reductions of non-conformal branes (the dVV formulation
was discussed for nonconformal branes in \cite{Sahakian:2000xx}) is of
the form
\begin{equation}
U (\Phi,\Psi) = e^{\gamma\Psi} {\tilde U}(\Phi)\ ,
\end{equation}
with \eg\ ${\tilde U}=-V_0\Phi+\frac{V_2}{\Phi^3}$ for 4-dim theories with
$z=1,\theta\neq 0$, as we have seen.
Assuming an ansatz for $W$, $W=e^{\frac{\gamma\Psi}{2}}\chi(\Phi)$
and substituting in \eqref{2dpotentialsrelation}, we get
\begin{equation}
\frac{d\chi^2}{d\Phi^2}=\frac{\gamma^2\chi^2}{2\Phi^2}-\tilde{U}\ .
\end{equation}
Integrating this equation, the general solution is
\begin{equation}
\chi^2=\chi_0(\Phi^2)^{\frac{\gamma^2}{2}} -(\Phi^2)^{\frac{\gamma^2}{2}}\int d\Phi^2
\tilde{U}(\Phi^2)^{\frac{-\gamma^2}{2}}\ .
\end{equation}
Then $\beta_{\Phi}$ can be written as
\begin{equation}
\beta_{\Phi}=\frac{W^2}{\Phi\frac{\partial W^2}{\partial\Phi^2}}=\frac{\chi^2}{\Phi\frac{d\chi^2}{d\Phi^2}}=\frac{\chi^2}{\Phi(\frac{\gamma^2\chi^2}{2\Phi^2}-\tilde{U})}\ .
\end{equation}
For arbitrary $\chi_0$ such that $\chi\vert_h\neq 0$ at the critical point, $\beta_{\Phi}$ becomes $\beta_{\Phi}\Big\vert_h=\frac{2\Phi_h}{\gamma^2}\neq 0$.

Let us consider the case when $\chi_0$ is chosen such that $\chi\vert_h=0$:
this makes the boundary potential vanish at the critical point, \ie\
$W\vert_h=0$, corresponding to zero energy as in the conformal case above.
To study this case, we expand $\tilde{U}$ around the critical point,
\begin{equation}
\tilde{U}=\Big(\frac{d\tilde{U}}{d\Phi^2}\Big\vert_h\Big) (\Phi^2-\Phi_h^2)
+ \cdots\ , \qquad \text{where}\quad \tilde{U}\vert_h=0\ ,
\end{equation}
and substitute the solution for $\chi$ in the above expression for
$\beta$-function to get
\begin{equation}
\beta_{\Phi}=\frac{2\Phi_h}{\gamma^2}\frac{\Big[\chi_0-(\Phi_h^2)^{1-\frac{\gamma^2}{2}}\Big(\frac{d\tilde{U}}{d\Phi^2}\Big\vert_h\Big)(\frac{\Phi^2}{2}-\Phi_h^2)\Big]}{\Big[\chi_0-(\Phi_h^2)^{1-\frac{\gamma^2}{2}}\Big(\frac{d\tilde{U}}{d\Phi^2}\Big\vert_h\Big)\Big(\frac{\Phi^2}{2}\Big(1-\frac{4}{\gamma^2}\Big)-\Phi_h^2(1-\frac{4}{\gamma^2}\Big)\Big)\Big]}\ .
\end{equation}
Choosing $\chi_0=-(\Phi_h^2)^{1-\frac{\gamma^2}{2}}\Big(\frac{d\tilde{U}}{d\Phi^2}\Big\vert_h\Big)\frac{\Phi_h^2}{2}$, which makes $\chi\vert_h=0$ (and so
$W\vert_h=0$), the above expression simplifies to
\begin{equation}
\beta_{\Phi}= {2\Phi_h\over \gamma^2-4}\ .
\end{equation}
For $W=e^{\frac{\gamma\Psi}{2}}\chi(\Phi)$, \eqref{2dbetapsi} and \eqref{2dbetaphi} give $\beta_{\Psi}=\frac{-2\gamma}{\Phi}\beta_{\Phi}$. We see that
both $\beta$-functions $\beta_{\Phi}$ and $\beta_{\Psi}$ do not vanish
at the $AdS_2$ critical point for any choice of $\chi_0$. This vindicates
the intuition that the $AdS_2$ critical point can consistently be placed
at the fixed point of an RG flow, but not at some intermediate point
along the flow.

\subsection{Examples}

\noindent\underline{$\beta$-function, $M2$-phase}:\ \ 
The effective 2-dim potential for 4-dim Einstein-Maxwell redux is
\begin{equation}\label{UM2}
U=-V_0\Phi+\frac{V_2}{\Phi^3}\ , \quad V_0=-2\Lambda_{(4)}=6\ , \quad V_2=2Q^2\ .
\end{equation}
Then for $W=W(\Phi)$, \eqref{2dpotentialsrelation} gives
\begin{equation}
W\frac{\partial W}{\partial\Phi}-V_0\Phi^2+\frac{V_2}{\Phi^2}=0\ , \qquad\ie\
\qquad \frac{\partial}{\partial\Phi}\Big(\frac{W^2}{2}\Big)=V_0\Phi^2-\frac{V_2}{\Phi^2}\ .
\end{equation}
Integrating this equation and imposing $W\vert_h=0$, we obtain
\begin{equation}
W=-\Big[\frac{2V_0\Phi^3}{3}+\frac{2V_2}{\Phi}+2\chi_0\Big]^{\frac{1}{2}}\ ,
\end{equation}
where the integration constant $\chi_0$ is fixed by our boundary
condition (\ref{WPhiconf}) to be $\chi_0=-8r_0^3$ using (\ref{UM2}).
The $\beta$-function using \eqref{2dbetaphi} is
\begin{equation}
\beta_{\Phi}=\frac{\Big[\frac{2V_0\Phi^3}{3}+\frac{2V_2}{\Phi}+2\chi_0\Big]}{\Big[V_0\Phi^2-\frac{V_2}{\Phi^2}\Big]}\ .
\end{equation}
As we approach the $AdS_2$ critical point placed in the $M2$ phase\
$\Phi_h=r_0 , \ Q^2=3r_0^4$,\ we see that $\beta_{\Phi}$ vanishes,
elucidating the general discussion above for conformal branes (note
that $\beta_{\Phi}$ diverges for arbitrary $\chi_0$ so the boundary
condition on $W$ is important).

\vspace{1mm}

\noindent\underline{$\beta$-function, $D2$-phase}:\ \ 
D2-branes after $S^6$-redux lead to a 4-dim hvLif theory with exponents
$z=1$, $\theta=-\frac{1}{3}$\,. The effective potential in the 2-dim
corresponding theory, again setting dimensionful parameters to unity
for convenience, is
\begin{equation}\label{UD2}
U=e^{\gamma\Psi}\Big(-V_0\Phi+\frac{V_2}{\Phi^3}\Big)\ , \qquad \gamma=-\frac{-1}{\sqrt{7}}\ , \quad V_0=\frac{70}{9}\ , \quad V_2=\frac{28}{9}Q^2\ .
\end{equation}
Assuming an ansatz\ $W=e^{\frac{\gamma\Psi}{2}}\chi(\Phi)$ for $W$ and
substituting in \eqref{2dpotentialsrelation} gives
\begin{equation}
\chi\frac{\partial\chi}{\partial\Phi}-\frac{\gamma^2\chi^2}{2\Phi}-V_0\Phi^2+\frac{V_2}{\Phi^2}=0\ ,
\end{equation}
whose solution gives
\begin{equation}
W=e^{\frac{\gamma\Psi}{2}}\chi(\Phi)=-e^{\frac{\gamma\Psi}{2}}\Big[\frac{49}{9}\Big(\frac{Q^2}{\Phi}+\Phi^3\Big)+\chi_0\Phi^{\frac{1}{7}}\Big]^{\frac{1}{2}}\ ,
\end{equation}
where the integration constant $\chi_0$ is again fixed by the boundary
condition $W\vert_h=0$\ (it will turn out that the precise value of
$\chi_0$ drops out in what follows).
The $\beta$-functions from \eqref{2dbetaphi}, \eqref{2dbetapsi} become
\begin{equation}\label{betaD2}
\beta_{\Phi}=\frac{2\Phi\Big[\frac{49}{9}(Q^2+\Phi^4)+\chi_0\Phi^{\frac{8}{7}}\Big]}{\Big[\frac{49}{9}(-Q^2+3\Phi^4)+\frac{\chi_0}{7}\Phi^{\frac{8}{7}}\Big]}\
\xrightarrow{h}\ 14 r_0^{\frac{7}{6}} \ , \qquad
\beta_{\Psi}=\frac{4}{\sqrt{7}}\frac{\Big[\frac{49}{9}(Q^2+\Phi^4)+\chi_0\Phi^{\frac{8}{7}}\Big]}{\Big[\frac{49}{9}(-Q^2+3\Phi^4)+\frac{\chi_0}{7}\Phi^{\frac{8}{7}}\Big]}\ \xrightarrow{h}\ 4\sqrt{7}\ ,
\end{equation}
where we have evaluated the $\beta$-functions at the $AdS_2$ critical point
placed in this $D2$-phase, which has
\be
\Phi_h=r_0^{\frac{7}{6}}\ , \quad e^{\frac{\gamma\Psi_h}{2}}=r_0^{-\frac{1}{6}}\ ,
\quad Q^2=\frac{5}{2}r_0^{\frac{14}{3}}\ \ \Rightarrow\ \
\Phi^4+Q^2=\frac{7}{2}r_0^{\frac{14}{3}}\ , \quad
3\Phi^4-Q^2=\frac{1}{2}r_0^{\frac{14}{3}}\ .
\ee
These nonvanishing $\beta$-functions imply that the theory is still
flowing at the $AdS_2$ critical point which thus is an inconsistency
and shows up as the massless scalar mode found previously: the $AdS_2$
horizon is only consistently placed within the true fixed point region
of the RG flow which is the above $M2$-phase in this D2-M2 phase
diagram.

\vspace{1mm}

\noindent\underline{$\beta$-function, $M5$-$D4$ phases}:\ \
We can likewise analyse the flow for the M5-D4 system: here the M5-$AdS_7$
phase (after reducing on the $S^4$) flows to the D4-supergravity phase
obtained by dimensional reduction on the M-theory 11th circle. The
$AdS_7$ phase has $z=1, \theta=0$ while the D4-phase is a 6-dim hvLif
theory with $z=1, \theta=-1$ and again the scalar leads to a massless
mode if the $AdS_2$ horizon is placed within this region. Here again,
the $\beta$-functions can be shown to vanish in the conformal
$M5$-phase but not in the $D4$-phase. To obtain extremal branes, we
add an additional $U(1)$ gauge field which provides charge: this gives
an Einstein-Maxwell or Einstein-Maxwell-scalar theory in the M5- and
D4-phases respectively.

The effective 2-dim potential for 7-dim Einstein-Maxwell redux is
\begin{equation}
U=-V_0\Phi^{\frac{2}{5}}+\frac{V_2}{\Phi^{\frac{18}{5}}}
=\frac{1}{\Phi^{\frac{3}{5}}}\Big(-V_0\Phi+\frac{V_2}{\Phi^3}\Big)\ ,
\quad V_0=-2\Lambda_{(7)}=30\ , \quad V_2=20\,Q^2\ .
\end{equation}
Then \eqref{2dpotentialsrelation} as for $M2$-branes gives
\begin{equation}
W=-\Big[\frac{5V_0\Phi^{\frac{12}{5}}}{6}+\frac{5V_2}{4\Phi^{\frac{8}{5}}}
-\frac{125}{2}r_0^6\Big]^{\frac{1}{2}}\ ,
\end{equation}
where an integration constant has again been fixed by the boundary
condition $W\vert_h=0$. Then the $\beta$-function using \eqref{2dbetaphi} is
\begin{equation}
\beta_{\Phi}=\frac{\Big[\frac{5V_0\Phi^{\frac{12}{5}}}{6}+\frac{5V_2}{4\Phi^{\frac{8}{5}}}-\frac{125}{2}r_0^6\Big]}{\Phi\Big[V_0\Phi^{\frac{2}{5}}-\frac{V_2}{\Phi^{\frac{13}{5}}}\Big]}\ ,
\end{equation}
which vanishes at the $AdS_2$ horizon\ 
$\Phi_h=r_0^{\frac{5}{2}} ,\ Q^2=\frac{3}{2}r_0^{10}$,\
if placed in the $M5$ phase.

Now, for the 6-dim hvLif theory with $z=1$, $\theta=-1$ from D4-redux,
the 2-dim effective potential is
\begin{equation}
U=e^{\gamma\Psi}\Big(-V_0\Phi^{\frac{1}{2}}+\frac{V_2}{\Phi^{\frac{7}{2}}}\Big)=e^{\gamma\Psi}\Big(-V_0\Phi+\frac{V_2}{\Phi^3}\Big)\frac{1}{\Phi^{\frac{1}{2}}}\ , \qquad \gamma=-\frac{-1}{\sqrt{10}}\ , \quad V_0=30\ , \quad V_2=20\,Q^2\ .
\end{equation}
As for the D2-case, taking an ansatz $W=e^{\frac{\gamma\Psi}{2}}\chi(\Phi)$
and using \eqref{2dpotentialsrelation} gives 
\begin{equation}
W=e^{\frac{\gamma\Psi}{2}}\chi(\Phi)=-e^{\frac{\gamma\Psi}{2}}\Big[\frac{25Q^2}{\Phi^{\frac{3}{2}}}+25\Phi^{\frac{5}{2}}+\chi_0\Phi^{\frac{1}{10}}\Big]^{\frac{1}{2}}\ ,
\end{equation}
where the precise value of the integration constant $\chi_0$ will again
not play any role. The $\beta$-functions using \eqref{2dbetaphi},
\eqref{2dbetapsi} are
\begin{equation}
\beta_{\Phi}=2\Phi\frac{\Big[\frac{25Q^2}{\Phi^{\frac{3}{2}}}+25\Phi^{\frac{5}{2}}+\chi_0\Phi^{\frac{1}{10}}\Big]^{\frac{1}{2}}}{\Big[\frac{-75Q^2}{2\Phi^{\frac{3}{2}}}+\frac{125}{2}\Phi^{\frac{5}{2}}+\frac{\chi_0}{10}\Phi^{\frac{1}{10}}\Big]}\ , \qquad \beta_{\Psi}=\frac{4}{\sqrt{10}}\frac{\Big[\frac{25Q^2}{\Phi^{\frac{3}{2}}}+25\Phi^{\frac{5}{2}}+\chi_0\Phi^{\frac{1}{10}}\Big]^{\frac{1}{2}}}{\Big[\frac{-75Q^2}{2\Phi^{\frac{3}{2}}}+\frac{125}{2}\Phi^{\frac{5}{2}}+\frac{\chi_0}{10}\Phi^{\frac{1}{10}}\Big]}\ .
\end{equation}
If the $AdS_2$ critical point is placed in the $D4$ phase, we require
\begin{equation}
\Phi_h=r_0^{\frac{5}{2}}\ , \quad Q^2=\frac{3}{2}r_0^{10}\ \ \Rightarrow\ \
\frac{25Q^2}{\Phi^{\frac{3}{2}}}+25\Phi^{\frac{5}{2}}=\frac{125}{2}r_0^{\frac{25}{4}}\ , \quad \frac{-75Q^2}{2\Phi^{\frac{3}{2}}}+\frac{125}{2}\Phi^{\frac{5}{2}}=\frac{25}{4}r_0^{\frac{25}{4}}\ ,
\end{equation}
giving\
$\beta_{\Phi}|_h\ra 20\,r_0^{\frac{5}{2}}$\ and\
$\beta_{\Psi}|_h\ra 4\sqrt{10}$\,.\
As in the D2-case (\ref{betaD2}), these nonvanishing $\beta$-functions
imply that it is inconsistent to place the $AdS_2$ critical point in the
D4-phase where the theory has a nontrivial RG flow.

It appears nontrivial to carry out this analysis of the flow equations
and $\beta$-functions for general potential $U(\Phi,\Psi^I)$ as
\eg\ for more general hvLif theories. Since the perturbation analysis
in \cite{Kolekar:2018sba} revealed a disconcerting massless mode only
for $z=1$ (which dovetails with our analysis here), it would appear
that there would be no problem for the $AdS_2$ throat to emerge in
general hvLif$_{z,\theta}$ theories. It would be interesting to explore
this further.

\section{Discussion}

We have formulated a version of the holographic renormalization group
flow for 2-dim dilaton-gravity-scalar theories arising from reductions
of higher dimensional extremal black branes, as in
\cite{Kolekar:2018sba}, thereby restricting to 2-dim flows that end at
an $AdS_2$ throat.  We have assumed that the transverse space is
sufficiently symmetric which then allows this formulation to be
insensitive to the higher dimensional branes being relativistic or
nonrelativistic.  Based on the null energy conditions, we have
proposed a holographic c-function in terms of the 2-dim dilaton and
given arguments for the corresponding c-theorem (subject to
appropriate boundary conditions on the ultraviolet theory): at the IR
$AdS_2$, this becomes the extremal black brane entropy. We have
discussed this c-function (essentially inherited from higher
dimensions) in detail for nonconformal branes compactified, and
compared with other c-functions. Finally, we have adapted the radial
Hamiltonian flow formulation of \cite{deBoer:1999tgo} to these 2-dim
theories: while this is not Wilsonian, it gives qualitative insight
into the flow equations and $\beta$-functions.

It would be interesting to understand how general such a holographic
RG flow is. For instance, since our formulation has crucially used the
sufficiently high symmetry of the transverse space, it is unclear if
this directly applies to other situations, involving \eg\ rotation
(see \eg\ \cite{Castro:2018ffi}). It is also important to note that
unlike a black hole which exhibits a gap, the branes we have
considered would contain additional low-lying modes: from our
analysis, it would seem that these do not change the essential flow
pattern, \ie\ the c-function does capture the relevant degrees of
freedom describing the effective 2-dim physics. This is additionally
corroborated by the fact that in the infrared it equals the extremal
entropy which is the number of available microstates.

The analysis adapting \cite{deBoer:1999tgo} was motivated by the fact
that the scalar perturbation mode in \cite{Kolekar:2018sba} about the
$AdS_2$ background was found to be massless for $z=1$ hvLif theories:
this includes the hvLif family arising from reductions of nonconformal
branes.  We have seen however that in this case the $\beta$-functions
do not vanish, whereas they do for reductions of the M2-$AdS_4$ phase
to $AdS_2$. This suggests that it is consistent for the $AdS_2$ throat
to emerge in a conformal phase of the higher dimensional theory (with
$AdS_D$ dual) but not consistent to have the $AdS_2$ critical point
lie within a region encoding nontrivial RG flow. This is exemplified
in the D2-M2 phase diagram and is consistent with our discussion of
the c-function in sec.~\ref{cfnD2M2}. This dVV formulation is not
Wilsonian, as discussed in the literature: it would be interesting to
adapt the Wilsonian formulations of
\cite{Heemskerk:2010hk,Faulkner:2010jy} to the 2-dim context: we hope
to report on this in the future.  Relatedly it would be interesting to
explore holographic renormalization \cite{Henningson:1998gx,de
  Haro:2000xn,Skenderis:2002wp} (\cite{Kanitscheider:2008kd} for
uncharged nonconformal branes) in this 2-dim context, perhaps building
on \cite{Cvetic:2016eiv}.

For the 2-dim theories in \cite{Kolekar:2018sba} arising from
compactification, the leading departures away from the IR $AdS_2$
critical point, described by Jackiw-Teitelboim theory, arise from the
leading linear term in the dilaton perturbation and are thus governed
by the Schwarzian derivative effective action.  The dilaton
fluctuation in (\ref{2d-fluctEOM}) has $m^2L^2=2$ and so corresponds
to an irrelevant operator with dimension $\Delta=2$\ (using
$\Delta={1\over 2}+\sqrt{{1\over 4}+m^2L^2}$ for a scalar mode
$\varphi$ of mass $m$ with equation of motion\
$\partial_+\partial_-\varphi+\frac{m^2L^2}{(x^+ - x^-)^2}\varphi=0$).
In light of the present work we note that some of the more general
2-dim dilaton-gravity-matter theories (\ref{2dimgsaction}) in the IR
$AdS_2$ region may contain fluctuation modes with masses\ $-{1\over
  4}\leq m_I^2L^2<2$\ corresponding to dual operators with dimension
$\Delta<2$. In such cases, the leading departures from the IR $AdS_2$
will presumably not be governed by the Schwarzian but some distinct
effective theory. It would be interesting to explore this further.

Our analysis here raises the question of understanding renormalization
group flow in boundary quantum mechanical theories, which could be
interpreted as flowing to lower energies (although not as spatial
coarse-graining). The discussions here on \eg\ nonconformal branes all
pertain to large $N$ (highly) supersymmetric theories (although fairly
complicated, since in the IR they are dual to the compactified
extremal black branes). Although we have not used this, it would seem
that the constraints from supersymmetry will be powerful in
1-dimension, just as in higher dimensions as is well-known. It would
be interesting to explore this.

Finally it is interesting to ask if the $2$-dim dilaton-gravity-scalar
theories of the general form (\ref{2dimgsaction}) we have considered
admit 2-dim de Sitter space $dS_2$ as solutions. For simplicity,
taking $\Psi^I=0$ and constant dilaton $\Phi$, the Einstein equations
and dilaton equation \eqref{2dimgseoms} require $U=0$ and
$\frac{\partial U}{\partial\Phi^2}=\mathcal{R}>0$ at the $dS_2$
critical point.  However this violates the condition \eqref{NEC2-UdU}
which we expect must hold if we take the potential $U$ as arising from
some higher dimensional reduction as we have discussed (implicitly
taking $U$ to have a leading term arising from a negative cosmological
constant as in known brane realizations followed by positive flux
contributions). Of course there are rolling (time-dependent) scalar
solutions, as \eg\ arises from reductions of $dS_4$\ (say with
Poincare metric $ds^2={R_{dS}^2\over\tau^2}(-d\tau^2+dw^2+dx_i^2)$)\,.
In 4-dim Einstein gravity with a positive cosmological constant
$\Lambda>0$, the 2-dim potential simply becomes $U=2\Lambda\Phi>0$,
and the 2-dim dilaton is $\Phi^2\sim {1\over\tau^2}$\,. The nature of
such solutions (even in this simple classical sense) might be
different from our $AdS_2$ discussions here and might be worth
exploring (see \eg\ \cite{Anninos:2017hhn}).

\vspace{7mm}

{\footnotesize \noindent {\bf Acknowledgements:}\ \ It is a pleasure
  to thank Micha Berkooz, Juan Maldacena and Ashoke Sen for
  discussions on \cite{Kolekar:2018sba}, and Ioannis Papadimitriou for
  some useful early correspondence. While this work was in progress,
  KK thanks the hospitality of NISER Bhubaneshwar during the ST4
  school, Jul 2018 and the organizers of the PiTP school ``From Qubits
  to spacetime'', IAS, Princeton, Jul 2018; KN thanks the organizers
  of ``Strings 2018'', Okinawa, and the longterm Yukawa memorial
  workshop ``New Frontiers in String Theory'', Yukawa Institute,
  Kyoto, Japan, Jun-Jul 2018. The research of KK and KN was supported
  in part by the International Centre for Theoretical Sciences (ICTS)
  during a visit for participating in the program - AdS/CFT at 20 and
  Beyond (Code: ICTS/adscft20/05). We both also thank the hospitality
  of the organizers of the Chennai Strings Meeting, IMSc Chennai, Oct
  2018.  This work is partially supported by a grant to CMI from the
  Infosys Foundation.  }

\appendix


\section{Effective potential in $D$-dim gravity-scalar
	theory}\label{DdimgsVeff}

We derive a formula for the effective potential $V$ in gravity scalar action \eqref{Ddimgsaction} starting from gravity-scalar action coupled to $U(1)$ gauge fields. Consider the action
\begin{equation}
S=\frac{1}{16\pi G_D}\int d^Dx \sqrt{-g^{(D)}} \Big(\mathcal{R}^{(D)}-\frac{h_{IJ}}{2}\partial_M\Psi^I\partial^M\Psi^J+V_0(\Psi)-\sum_{i=1}^{n}\frac{Z_i(\Psi)}{4}F_i^2\Big)\ ,
\end{equation}
where $V_0(\Psi)$ is the scalar potential, $Z_i(\Psi)$ are $\Psi$ dependent couplings and $F_i^2=F_{i\,MN}F_i^{MN}$. The Einstein's equations are
\begin{eqnarray}
&& \mathcal{R}^{(D)}_{MN}-\frac{g^{(D)}_{MN}}{2}\mathcal{R}^{(D)}=\frac{h_{IJ}}{2}\Big(\partial_M\Psi^I\partial_N\Psi^J-\frac{g^{(D)}_{MN}}{2}\partial_P\Psi^I\partial^P\Psi^J\Big)+\frac{g^{(D)}_{MN}}{2}V_0 \nonumber \\
&& \hspace{4cm} +\sum_{i}^{n}\frac{Z_i}{2}\Big(F_{i\,MP}F_{i\,N}^P-\frac{g^{(D)}_{MN}F_i^2}{4}\Big)\ . \label{DdimEMsEE}
\end{eqnarray}
Taking electric ansatz for all gauge fields and solving the Maxwell's equations
\begin{equation}
\partial_M(\sqrt{-g^{(D)}}Z_i F_i^{MN})=0 \qquad \implies \quad F_i^{tr}=\frac{\tilde{c}_i}{\sqrt{-g^{(D)}}Z_i}\ , \qquad F_i^2=-\frac{2\,\tilde{c}_i^2}{(g^{(D)}_{xx})^{D-2}Z_i^2}\ ,
\end{equation}
where $\tilde{c}_i$ is constant and we have restricted to a class of metrics with $g^{(D)}_{ij}=\delta_{ij}g^{(D)}_{xx}$, consistent with the reduction ansatz \eqref{DdimKKmetric}. Substituting $F_i^{tr}$ in \eqref{DdimEMsEE}, the $t-r$ and $xx$ components become
\begin{equation}\label{DdimEMsEEtr}
G^{(D)}_{\mu\nu}=\frac{h_{IJ}}{2}\Big(\partial_{\mu}\Psi^I\partial_{\nu}\Psi^J-\frac{g^{(D)}_{\mu\nu}}{2}\partial_P\Psi^I\partial^P\Psi^J\Big)+\frac{g^{(D)}_{\mu\nu}}{2}\Big(V_0 -\sum_{i=1}^{n}\frac{\tilde{c}_i^2}{2(g^{(D)}_{xx})^{D-2}Z_i}\Big)\ ,
\end{equation}
\begin{equation}\label{DdimEMsEExx}
G^{(D)}_{xx}=\frac{h_{IJ}}{2}\Big(\partial_x\Psi^I\partial_x\Psi^J-\frac{g^{(D)}_{xx}}{2}\partial_P\Psi^I\partial^P\Psi^J\Big)+\frac{g^{(D)}_{xx}}{2}V_0 +\frac{g^{(D)}_{xx}}{4}\frac{\tilde{c}_i^2}{(g^{(D)}_{xx})^{D-2}Z_i}\ ,
\end{equation}
where $G^{(D)}_{MN}=\mathcal{R}^{(D)}_{MN}-\frac{g^{(D)}_{MN}}{2}\mathcal{R}^{(D)}$ is the Einstein tensor. These Einstein equations can be derived from an equivalent gravity-scalar action \eqref{Ddimgsaction}, with the effective potential defined as
\begin{equation}\label{Veff}
V(\Psi^I,g^{(D)}_{xx}) = -V_0(\Phi) +\sum_{i=1}^{n}\frac{V_i(\Phi)}{(g^{(D)}_{xx})^{D-2}}\ ,
\end{equation}
where $V_i(\Psi)\equiv \frac{\tilde{c}_i^2}{2\,Z_i(\Psi)}$.

\section{Radial Lagrangian \eqref{2dradialL} from dimensional
	reduction}\label{2dradialLderivation}

Consider the gravity scalar action in $D$-dimensions\\
\begin{equation}\label{DdimgsactionwithGH}
S=\frac{1}{16\pi G_D}\Big[\int d^Dx\sqrt{-g^{(D)}}\Big(\mathcal{R}^{(D)}-\frac{h_{IJ}}{2}\partial_M\Psi^I\partial^M\Psi^J-V \Big) +\int d^{D-1}x\sqrt{-\gamma^{(D-1)}}2K^{(D-1)}\Big],
\end{equation}
where $\gamma^{(D-1)}_{ab}$ is the induced metric and $K^{(D-1)}$ is the extrinsic curvature on the $(D-1)$-dimensional boundary. Foliating the spacetime into surfaces of constant $r$,
\begin{equation}\label{Ddimmetricrfoliation}
ds^2=g^{(D)}_{MN}dx^M dx^N=(\tilde{N}^2+\gamma^{(D-1)}_{ab}N^a N^b)dr^2 +2\gamma^{(D-1)}_{ab}N^a dx^b dr +\gamma^{(D-1)}_{ab}dx^a dx^b\ ,
\end{equation}
the $D$-dim Ricci scalar decomposes as
\begin{equation}\label{DdimRiccidecomp}
\mathcal{R}^{(D)}=\,^{(D-1)}\mathcal{R}+(K^{(D-1)})^2-K^{(D-1)}_{ab}K^{(D-1)\,ab} -2\nabla_A(\tilde{n}^A K^{(D-1)})+2\nabla_A(\tilde{n}^B\nabla_B \tilde{n}^A)\ ,
\end{equation}
where the indices $M,N$ take values $(t,r,x^i)$ and $a,b$ take values $(t,x^i)$ for $i=1,\dots,D-2$. $^{(D-1)}\mathcal{R}$ is the Ricci scalar of the $(D-1)$-dim boundary and $\tilde{n}^A$ is the unit normal to the boundary. The total derivative terms above can be written as
\begin{eqnarray}
&& \int d^Dx \sqrt{-g^{(D)}}[-2\nabla_A(\tilde{n}^A K^{(D-1)})+2\nabla_A(\tilde{n}^B\nabla_B \tilde{n}^A)] \nonumber \\
&& \quad =\int d^{D-1}x\sqrt{-\gamma^{(D-1)}}[-2K^{(D-1)}\tilde{n}^A\tilde{n}_A +\tilde{n}_A\tilde{n}^B\nabla_B \tilde{n}^A] =-\int d^{D-1}x\sqrt{-\gamma^{(D-1)}}2K^{(D-1)}, \nonumber
\end{eqnarray}
where we have used $\tilde{n}^A\tilde{n}_A=1$ and $\tilde{n}_A\nabla_B\tilde{n}^A=0$. This boundary term coming from the total derivative terms in \eqref{DdimRiccidecomp} cancels the Gibbons-Hawking term in \eqref{DdimgsactionwithGH}. Then the radial Lagrangian on the $r=constant$ boundary can be written as
\begin{eqnarray}
&& L=\frac{1}{16\pi G_D}\int d^{D-1}x \sqrt{-\gamma^{(D-1)}}\tilde{N}\Big(\,^{(D-1)}\mathcal{R}+(K^{(D-1)})^2-K^{(D-1)}_{ab}K^{(D-1)\,ab} \nonumber \\
&& \hspace{7cm} -\frac{h_{IJ}}{2}\partial_M\Psi^I\partial^M\Psi^J-V\Big)\ , \label{DdimradialL}
\end{eqnarray}
where the extrinsic curvature is
\begin{eqnarray}
&& K^{(D-1)}_{ab}=\frac{1}{2\tilde{N}}\Big(\partial_r\gamma^{(D-1)}_{ab}-D^{(D-1)}_a N^{(D-1)}_b -D^{(D-1)}_b N^{(D-1)}_a\Big)\ , \qquad N^{(D-1)}_a\equiv\gamma^{(D-1)}_{ab}N^b\ , \nonumber \\
&& D^{(D-1)}_a N^{(D-1)}_b=\partial_a N^{(D-1)}_b -\Gamma^{(D-1)\,c}_{ab}N^{(D-1)}_c\ .
\end{eqnarray}

\noindent \textbf{Radial decomposition of $D$-dim metric in the KK
	reduction form:}\\
Expanding the $D$-dim metric \eqref{Ddimmetricrfoliation}, into $2$-dim $(t,r)$ and transverse components
\begin{eqnarray}
&& ds^2=[(\tilde{N}^2+\gamma^{(D-1)}_{ab}N^a N^b)dr^2 +2\gamma^{(D-1)}_{tt}N^t dt dr+2\gamma^{(D-1)}_{ti}N^i dt dr +\gamma^{(D-1)}_{tt}dt^2] \nonumber \\
&& \qquad\quad +\gamma^{(D-1)}_{ij}dx^i dx^j +[2\gamma^{(D-1)}_{ti}N^t dx^i dr +2\gamma^{(D-1)}_{ij}N^i dx^j dr +2\gamma^{(D-1)}_{ti}dx^i dt]\ .
\end{eqnarray}
Imposing the Kaluza-Klein ansatz for the dimensional reduction on $T^{D-2}$, i.e.,
\begin{equation}
ds^2=g^{(2)}_{\mu\nu}dx^{\mu}dx^{\nu}+\Phi^{\frac{4}{D-2}}\sum_{i=1}^{D-2}dx_i^2\ ,
\qquad\qquad g_{xx}^{(D)}\equiv \Phi^{\frac{4}{D-2}}\ .
\end{equation}
where $g^{(2)}_{\mu\nu}$, $\Phi$ depend only on the $2$-dim coordinates $(t,r)$, we get
\begin{equation}
\gamma^{(D-1)}_{ij}=\Phi^\frac{4}{D-2}\delta_{ij}\ , \quad \gamma^{(D-1)}_{ti}=0\ , \quad N^i=0\ , \quad N^{(D-1)}_i=0\ , \quad \forall\ i,j=1,\dots,D-1
\end{equation}
and the components of the $2$-dim metric and its inverse are
\begin{eqnarray}
&& g^{(2)}_{rr}=g^{(D)}_{rr}=\tilde{N}^2+\gamma^{(D-1)}_{tt}(N^t)^2\ , \qquad g^{(2)}_{tr}=g^{(D)}_{tr}=\gamma^{(D-1)}_{tt}N^t\ , \qquad g^{(2)}_{tt}=g^{(D)}_{tt}=\gamma^{(D-1)}_{tt}, \nonumber \\
&& g^{(2)\,rr}=\frac{1}{\tilde{N}^2}\ , \qquad g^{(2)\,tr}=-\frac{N^t}{\tilde{N}^2}\ , \qquad g^{(2)\,tt}=\frac{1}{\gamma^{(D-1)}_{tt}}+\frac{(N^t)^2}{\tilde{N}^2}\ .
\end{eqnarray}
\smallskip

\noindent\textbf{Reduction of the radial Lagrangian \eqref{DdimradialL}:}\\
The induced metric on the $(D-1)$-dim boundary can be written as
\begin{equation}
ds_{(D-1)}^2=\gamma^{(D-1)}_{ab}dx^a dx^b=\gamma^{(D-1)}_{tt}dt^2+\gamma^{(D-1)}_{ij}dx^i dx^j=\gamma^{(D-1)}_{tt}dt^2+ \Phi^\frac{4}{D-2}\sum_{i=1}^{D-2}dx_i^2\ .
\end{equation}
The Ricci scalar becomes
\begin{equation}\label{3dbdyricci}
^{(D-1)}\mathcal{R}=\frac{\partial_t\gamma^{(D-1)}_{tt}\partial_t\Phi^2}{(\gamma^{(D-1)}_{tt})^2\,\Phi^2}+\frac{(D-3)}{(D-2)}\frac{(\partial_t\Phi^2)^2}{\gamma^{(D-1)}_{tt}\Phi^4}-\frac{2\,\partial_t^2\Phi^2}{\gamma^{(D-1)}_{tt}\Phi^2}\ .
\end{equation}
The components of the extrinsic curvature are
\begin{eqnarray}
&& K^{(D-1)}_{tt}=\frac{1}{2\tilde{N}}\Big(\partial_r\gamma^{(D-1)}_{tt}-2D^{(D-1)}_t N^{(D-1)}_t\Big)=\frac{1}{2\tilde{N}}\Big(\partial_r\gamma^{(1)}_{tt}-2D^{(1)}_t N^{(1)}_t\Big)=K^{(1)}_{tt}\ , \nonumber \\
&& K^{(D-1)}_{ti}=\frac{1}{2\tilde{N}}\Big(\partial_r\gamma^{(D-1)}_{ti}-D^{(D-1)}_t N^{(D-1)}_i -D^{(D-1)}_i N^{(D-1)}_t\Big)=0\ , \\
&& K^{(D-1)}_{ij}=\frac{1}{2\tilde{N}}\Big(\partial_r\gamma^{(D-1)}_{ij}-D^{(D-1)}_i N^{(D-1)}_j -D^{(D-1)}_j N^{(D-1)}_i\Big)=\delta_{ij}\frac{\Phi^{\frac{4}{D-2}-2}}{(D-2)}\tilde{n}^{\mu}\partial_{\mu}\Phi^2\ , \nonumber
\end{eqnarray}
where we have used
\begin{eqnarray}
&& \gamma^{(1)}_{tt}=\gamma^{(D-1)}_{tt}\ , \qquad N^{(1)}_t=\gamma^{(1)}_{tt}N^t=\gamma^{(D-1)}_{tt}N^t=N^{(D-1)}_t\ , \qquad \Gamma^{(D-1)\,t}_{it}=0\ , \nonumber \\
&& \Gamma^{(1)\,t}_{tt}=\Gamma^{(D-1)\,t}_{tt}\ , \qquad \tilde{n}_r=\tilde{N}\ , \quad \tilde{n}_t=0\ , \quad \tilde{n}^{\mu}=g^{(2)\,\mu\nu}\tilde{n}_{\nu}\ , \nonumber \\
&&  D^{(D-1)}_t N^{(D-1)}_t=\partial_t N^{(D-1)}_t-\Gamma^{(D-1)\,t}_{tt}N^{(D-1)}_t=\partial_t N^{(1)}_t-\Gamma^{(1)\,t}_{tt}N^{(1)}_t=D^{(1)}_t N^{(1)}_t,
\end{eqnarray}
where $\gamma^{(1)}_{tt}$ is the induced metric and $K^{(1)}_{tt}$ is the extrinsic curvature on the $1$-dim boundary and $\tilde{n}^{\mu}$ is the outward unit normal to the $1$-dim boundary. Then we can compute
\begin{eqnarray}
K^{(D-1)\,ab}K^{(D-1)}_{ab}&=&\gamma^{(D-1)\,ac}\gamma^{(D-1)\,bd}K^{(D-1)}_{ab}K^{(D-1)}_{cd} \nonumber \\
&=&(\gamma^{(D-1)\,tt})^2(K^{(D-1)}_{tt})^2+\gamma^{(D-1)\,ik}\gamma^{(D-1)\,jl}K^{(D-1)}_{ij}K^{(D-1)}_{kl} \nonumber \\
&=&K^{(1)\,tt}K^{(1)}_{tt}+\frac{(\tilde{n}^{\mu}\partial_{\mu}\Phi^2)^2}{(D-2)\Phi^4}
\end{eqnarray}
and
\begin{equation}
K^{(D-1)}=\gamma^{(D-1)\,ab}K^{(D-1)}_{ab}=\gamma^{(D-1)\,tt}K^{(D-1)}_{tt}+\gamma^{(D-1)\,ij}K^{(D-1)}_{ij}=K^{(1)}+\frac{\tilde{n}^{\mu}\partial_{\mu}\Phi^2}{\Phi^2}\ ,
\end{equation}
where $K^{(1)}=\gamma^{(1)\,tt}K^{(1)}_{tt}$. Substituting these in \eqref{DdimradialL}, the radial Lagrangian becomes
\begin{eqnarray}
L&=&\frac{1}{16\pi G_2}\int dt \sqrt{-\gamma^{(1)}_{tt}}\Phi^2\tilde{N}\Big[\,^{(D-1)}\mathcal{R}+\Big(K^{(1)}+\frac{\tilde{n}^{\mu}\partial_{\mu}\Phi^2}{\Phi^2}\Big)^2-K^{(1)\,tt}K^{(1)}_{tt} \nonumber \\
&& \hspace{6cm} -\frac{(\tilde{n}^{\mu}\partial_{\mu}\Phi^2)^2}{(D-2)\Phi^4} -\frac{h_{IJ}}{2}g^{(2)\,\mu\nu}\partial_{\mu}\Psi^I\partial_{\nu}\Psi^J-V\Big] \nonumber \\
&=&\frac{1}{16\pi G_2}\int dt \sqrt{-\gamma^{(1)}_{tt}}\Phi^2\tilde{N}\Big[\,^{(D-1)}\mathcal{R} +2K^{(1)}\frac{\tilde{n}^{\mu}\partial_{\mu}\Phi^2}{\Phi^2} +\frac{(D-3)}{(D-2)}\frac{(\tilde{n}^{\mu}\partial_{\mu}\Phi^2)^2}{\Phi^4} \nonumber \\
&& \hspace{6cm} -\frac{h_{IJ}}{2}g^{(2)\,\mu\nu}\partial_{\mu}\Psi^I\partial_{\nu}\Psi^J-V\Big]\ , \label{2dradialLnoWeyl}
\end{eqnarray}
where we have used $K^{(1)\,tt}K^{(1)}_{tt}=(K^{(1)})^2$.

\noindent\underline{Weyl transformation:} Performing a Weyl transformation on the $2$-dim bulk metric, $g_{\mu\nu}=\Phi^{\frac{2(D-3)}{(D-2)}}g^{(2)}_{\mu\nu}$ which induces a Weyl transformation on the $1$-dim boundary metric $\gamma_{tt}=\Phi^{\frac{2(D-3)}{(D-2)}}\gamma^{(1)}_{tt}$, we get
\begin{eqnarray}
ds^2=g_{\mu\nu}dx^{\mu}dx^{\nu}&=&\Big(\Phi^{\frac{2(D-3)}{(D-2)}}\tilde{N}^2+\gamma_{tt}(N^t)^2\Big)dr^2+2\gamma_{tt}N^t dt dr +\gamma_{tt}dt^2 \nonumber \\
&\equiv&(N^2+\gamma_{tt}(N^t)^2)dr^2+2\gamma_{tt}N^t dt dr +\gamma_{tt}dt^2\ ,
\end{eqnarray}
which is same as \eqref{2dmetricfoliation} with $N\equiv\Phi^{\frac{D-3}{D-2}}\tilde{N}$. Under the Weyl transformation, we have
\begin{eqnarray}
&& \Gamma^{(1)\,t}_{tt}=\Gamma^t_{tt}-\frac{(D-3)}{(D-2)}\frac{\partial_t\Phi}{\Phi}\ , \qquad \Gamma^t_{tt}=\frac{\partial_t\gamma_{tt}}{2\gamma_{tt}}\ , \qquad N_t=\gamma_{tt}N^t\ , \nonumber \\
&& D^{(1)}_t N^{(1)}_t=\partial_t(\gamma^{(1)}_{tt}N^t)-\Gamma^{(1)\,t}_{tt}\gamma^{(1)}_{tt}N^t=\Phi^{\frac{-2(D-3)}{(D-2)}}\Big(D_tN_t-\frac{(D-3)}{(D-2)}\frac{N_t\partial_t\Phi}{\Phi^2}\Big)\ .
\end{eqnarray}
The Ricci scalar \eqref{3dbdyricci} can be written covariantly as
\begin{equation}
^{(D-1)}\mathcal{R}=-2\Phi^{\frac{2(D-3)}{(D-2)}-2}\Box_t\Phi^2\ ; \quad \Box_t\Phi^2=\gamma^{tt}D_t D_t\Phi^2=\frac{\partial_t(\sqrt{-\gamma_{tt}}\gamma^{tt}\partial_t\Phi^2)}{\sqrt{-\gamma_{tt}}}=\Big[\frac{\partial_t^2\Phi^2}{\gamma_{tt}}-\frac{\partial_t\gamma_{tt}\partial_t\Phi^2}{2\gamma_{tt}^2}\Big].
\end{equation}
The extrinsic curvature becomes
\begin{eqnarray}
&& K^{(1)}_{tt}=\frac{1}{2\tilde{N}}\Big(\partial_r\gamma^{(1)}_{tt}-2D^{(1)}_t N^{(1)}_t\Big)=\Phi^{-\frac{(D-3)}{(D-2)}}\Big(K_{tt}-\frac{(D-3)}{2(D-2)}\frac{\gamma_{tt}}{\Phi^2}n^{\mu}\partial_{\mu}\Phi^2\Big)\ , \\
&& K^{(1)}=\gamma^{(1)\,tt}K^{(1)}_{tt}=\Phi^{\frac{2(D-3)}{(D-2)}}\gamma^{tt}K^{(1)}_{tt}=\Phi^{\frac{(D-3)}{(D-2)}}\Big(K-\frac{(D-3)}{2(D-2)}\frac{n^{\mu}\partial_{\mu}\Phi^2}{\Phi^2}\Big)\ ,
\end{eqnarray}
where $K_{tt}=\frac{1}{2N}(\partial_r\gamma_{tt}-2D_tN_t)$, $K=\gamma^{tt}K_{tt}$, $n_r=N=\Phi^{\frac{D-3}{D-2}}\tilde{N}=\Phi^{\frac{D-3}{D-2}}\tilde{n}_r$ and $n^{\mu}\partial_{\mu}\Phi^2=g^{\mu\nu}n_{\mu}\partial_{\nu}\Phi^2=\Phi^{\frac{D-3}{D-2}}\tilde{n}^{\mu}\partial_{\mu}\Phi^2$. Substituting these expressions in \eqref{2dradialLnoWeyl}, we get
\begin{eqnarray}
L = \frac{1}{16\pi G_2}\int dt \sqrt{-\gamma}N\Big[-2\Box_t\Phi^2 +2K\,n^{\mu}\partial_{\mu}\Phi^{2}-\frac{\Phi^2 h_{IJ}}{2}\partial_{\mu}\Psi^I\partial^{\mu}\Psi^J-V\Phi^{\frac{2}{D-2}}\Big]\ .
\end{eqnarray}
Simplifying further using $n^{\mu}\partial_{\mu}\Phi^2=\frac{1}{N}(\partial_r\Phi^2-N^t\partial_t\Phi^2)$ and defining $U\equiv \Phi^{\frac{2}{D-2}} V$, we obtain
(\ref{2dradialL}).

\section{The $z=1, \theta\neq 0$ hvLif family}\label{z=1nonzerotheta}

Setting $z=1$, $\theta\neq 0$ in the $4$-dim Einstein-Maxwell-scalar
action \eqref{chbb4daction} and the charged hvLif solution
\eqref{chargedhvLifmetric}, \eqref{4dchargedhvLiffields} gives
$F_{1\,MN}=0$ and we get Einstein-scalar theory coupled to an $U(1)$
gauge field $A_{2\,M}$, with
\begin{equation}
V_0=\frac{(3-\theta)(2-\theta)\,e^{-\gamma\Psi_0}}{R^{2-2\theta}\,r_{hv}^{2\theta}}
  \ , \qquad \gamma=\frac{\theta}{\sqrt{(2-\theta)(-\theta)}}\ ,
\qquad \lambda_2=\sqrt{\frac{-\theta}{2-\theta}} \ \ .
\end{equation}
Note that the energy conditions \eqref{chbbzthetarange} simplify in this
case to give
\be
(2-\theta)\geq 0\ ,\qquad -\theta\geq 0 \qquad \Rightarrow\qquad
\gamma=-\lambda_2\ .
\ee
Substituting the gauge field solution in terms of the scalar field and
the metric component $g^{(4)}_{xx}$, gives an effective gravity-scalar
theory \eqref{Ddimgsaction} (in $4$-dim with one scalar field) with
an effective potential 
\bea\label{VeffPsiAction}
V_{eff}=-\frac{(3-\theta)(2-\theta)}{R^{2-2\theta}r_{hv}^{2\theta}}e^{\gamma(\Psi-\Psi_0)}\
+\ \frac{1}{(g^{(4)}_{xx})^2} \frac{(2-\theta)(1-\theta)Q^2 R^{-6+2\theta}}{e^{\lambda_2(\Psi-\Psi_0)}}\ .
\eea
Using the extremality condition $Q^2 = ({3-\theta\over 1-\theta})\, r_0^{2(2-\theta)}$ and $\gamma=-\lambda_2$ simplifies the effective potential
(\ref{chbb2deffactionWeyl}) which is of the form\ (\ref{UD2}) with
$U=\Phi V_{eff}$ and $\Phi=g_{xx}^{(4)}$.\ 
The factors of $\Phi$ arise, as reviewed in Sec.-\ref{Review},
from the $T^2$-compactification followed by a Weyl transformation of
the 2-dim metric. Thus the potential $U(\Phi,\Psi)$ has factorized in
this case: the piece inside the brackets is structurally similar to
that for the reduction of the M2-$AdS_4$ case, with an overall $\Psi$
factor dressing outside.

The linear fluctuations $\phi$, $\psi$, $\Omega$ to the dilaton,
scalar field and the metric respectively are governed by the quadratic
action (Sec.-3.2.1 in \cite{Kolekar:2018sba}),
which gives the linearized equation $\partial_+\partial_-\zeta=0$\ for
the fluctuation, $\zeta = \psi-\frac{2}{\sqrt{2-\theta}}\,
\frac{L^2(3-\theta)(2-\theta)}{r_0^{2-2\theta}r_{hv}^{2\theta}}\, \phi$.
Thus the $\zeta$ scalar is massless at linear order.



\end{document}